\documentclass[journal,comsoc]{IEEEtran}

\label{begin}

\usepackage[T1]{fontenc}

\usepackage{graphicx}
\usepackage{cite}
\usepackage{amsmath,amssymb,amsfonts,bm}
\usepackage{algorithmic}
\usepackage{setspace}
\usepackage{textcomp}

\usepackage{enumitem}
\setitemize[1]{itemsep=0pt,partopsep=0pt,parsep=\parskip,topsep=0pt,leftmargin=1em}

\usepackage[ruled,linesnumbered,nofillcomment]{algorithm2e} 
\usepackage{ascii}
\usepackage[bookmarks=false]{hyperref}

\hypersetup{draft}

\usepackage{threeparttable}
\usepackage{multicol}  
\usepackage{multirow}
\usepackage{arydshln}
\usepackage{makecell} 
\usepackage{amsfonts}
\usepackage{pifont}
\usepackage{booktabs}
\usepackage{colortbl}
\usepackage{ragged2e}
\usepackage[]{microtype}

\usepackage{tikz}
\newcommand*\emptycirc[1][1ex]{\tikz\draw[thick] (0,0) circle (#1);} 
\newcommand*\halfcirc[1][1ex]{%
	\begin{tikzpicture}
	\draw[fill] (0,0)-- (90:#1) arc (90:270:#1) -- cycle ;
	\draw[thick] (0,0) circle (#1);
	\end{tikzpicture}}
\newcommand*\fullcirc[1][1ex]{\tikz\fill (0,0) circle (#1);}

\usepackage{subcaption}
\ifCLASSINFOpdf
\else
\fi
\usepackage{amsmath}
\usepackage[cmintegrals]{newtxmath}
\begin{document}

		\onecolumn
		\copyright 2021 IEEE. Personal use of this material is permitted. Permission from IEEE must be obtained for all
		other uses, in any current or future media, including reprinting/republishing this material for advertising
		or promotional purposes, creating new collective works, for resale or redistribution to servers or lists, or
		reuse of any copyrighted component of this work in other works.

	\newpage 
	\twocolumn
	
	\title{Evaluating and Improving Adversarial Robustness of Machine Learning-Based Network \\ Intrusion Detectors}
	
	\author{Dongqi~Han,
		Zhiliang~Wang,~\IEEEmembership{Member,~IEEE,}
		Ying~Zhong,
		Wenqi~Chen,
		Jiahai~Yang,~\IEEEmembership{Senior Member,~IEEE,}
		Shuqiang~Lu,
		Xingang~Shi, ~\IEEEmembership{Member,~IEEE,}
		and~Xia~Yin, ~\IEEEmembership{Senior Member,~IEEE} 
	\thanks{This work was supported in part by the National Key Research and Development Program of China under Grant 2018YFB1800205.(\emph{Corresponding author:Zhiliang~Wang}).}
	\thanks{Dongqi Han, Zhiliang Wang, Ying Zhong, Wenqi Chen, Jiahai Yang, and Xingang Shi are with the Institute for Network Sciences and Cyberspace, Tsinghua University, Beijing 100084, China (e-mail: handq19@mails.tsinghua.edu.cn, wzl@cernet.edu.cn, zhongy18@mails.tsinghua.edu.cn, chenwq19@mails.tsinghua.edu.cn;  yang@cernet.edu.cn, shixg@cernet.edu.cn).}
	\thanks{Shuqiang Lu and Xia Yin are with the Department of Computer Science, Tsinghua University, Beijing 100084, China (e-mail: lusq18@mails.tsinghua.edu.cn; yxia@csnet1.cs.tsinghua.edu.cn).}}

	\maketitle
	
	\begin{abstract}
		Machine learning (ML), especially deep learning (DL) techniques have been increasingly used in anomaly-based network intrusion detection systems (NIDS).
		However, ML/DL has shown to be extremely vulnerable to adversarial attacks, especially in such security-sensitive systems. Many adversarial attacks have been proposed to evaluate the robustness of ML-based NIDSs. Unfortunately, existing attacks mostly focused on \emph{feature-space} and/or white-box attacks, which make impractical assumptions in real-world scenarios, leaving the study on practical gray/black-box attacks largely unexplored.
		
		To bridge this gap, we conduct the first systematic study of the gray/black-box \emph{traffic-space} adversarial attacks to evaluate the robustness of ML-based NIDSs.
		Our work outperforms previous ones in the following aspects:
		(\romannumeral1) \emph{practical}---the proposed attack can automatically mutate original traffic with extremely limited knowledge and affordable overhead while preserving its functionality;
		(\romannumeral2) \emph{generic}---the proposed attack is effective for evaluating the robustness of various NIDSs using diverse ML/DL models and non-payload-based features;
		(\romannumeral3) \emph{explainable}---we propose an explanation method for the fragile robustness of  ML-based NIDSs. Based on this, we also propose a defense scheme against adversarial attacks to improve system robustness.
		We extensively evaluate the robustness of various NIDSs using diverse feature sets and ML/DL models. 
		Experimental results show our attack is effective (e.g., >97\% evasion rate in half cases for \texttt{Kitsune}, a state-of-the-art NIDS) with affordable execution cost and the proposed defense method can effectively mitigate such attacks (evasion rate is reduced by >50\% in most cases).
	\end{abstract}
	
	\begin{IEEEkeywords}
		Network anomaly detection, network intrusion detection systems,  adversarial machine learning, machine learning security, evasion attack.
	\end{IEEEkeywords}
	
	%
	\IEEEpeerreviewmaketitle

	\section{Introduction}
	\label{sec1}
	
	\IEEEPARstart{N}{etwork} intrusion detection systems (NIDS) play a critical role in detecting malicious activities in networks. 
	Based on the detection mechanism, NIDSs can be generally classified into two types \cite{garcia2009anomaly}.
	\emph{Signature-based} NIDSs have traditionally been used to detect known malicious traffic, but are unable to detect new attacks or advanced variants.
	By contrast, \emph{anomaly-based} NIDSs, with machine learning (ML) and recently deep learning (DL) techniques, are receiving more and more attention due to the generalization ability to detect new and unknown attacks \cite{hodo_shallow_2017,gulghane2019survey,mirsky_kitsune:_2018,zhong2020helad,doriguzzi2020lucid,xu2020method,ferrag2020deep}. 
	
	While intrigued by the great promise and performance, ML and DL models have shown to be extremely vulnerable to \emph{adversarial attacks}---well-designed small changes of input can induce significant changes in the output of ML models\cite{barreno_security_2010,tygar_adversarial_2011,szegedy_intriguing_2014}.
	To evaluate the robustness of ML-based systems, the widely-used approach is constructing adversarial attacks to demonstrate the upper bound of the robustness \cite{carlini_towards_2017}.
	Such attacks must sufficiently reflect the adversary's ability in practical settings, otherwise will be useless for real-world systems.

	Several adversarial attacks have been proposed for ML-based systems in other domains such as computer vision \cite{goodfellow_explaining_2015,carlini_towards_2017, li_stealthy_2019}, natural language processing \cite{li_textbugger:_2019}, and malware detection \cite{laskov_practical_2014,xu_automatically_2016, hu_generating_2017, chen2020training}. However, these methods are ill-suited for network intrusion detection for two reasons. Firstly, such attacks are modifying feature vectors (we refer these to \emph{feature-space} attacks) instead of real input space (namely raw traffic for NIDSs). This is because in other domains the feature mapping from input space to feature space is reversible or can be easily formalized as differentiable functions.
	In contrast, feature extraction from network traffic to features in NIDS is neither invertible nor differentiable.
	As a result, feature-space attacks for NIDS are impractical in practice since adversarial traffic is difficult to derive from features. Secondly, we need to ensure that there is no communication violation or compromise of maliciousness when modifying malicious traffic, which is not a problem for other non-security domains.                          
	
	As for \emph{traffic-space} attacks (i.e., directly modifying traffic) against ML-based NIDSs, existing studies can be divided into three categories. Firstly, evasion attacks on traditional NIDSs have been well-studied \cite{ptacek_insertion_1998,handley_network_2001,cheng_evasion_2012}, but they are only useful for signature-based NIDSs without learning models. 
	Secondly, several traffic obfuscation and mutation methods have been proposed for evading traffic analysis based detection \cite{stinson_towards_2008, wright2009traffic, homoliak_improving_2018}. However, they are simply mimicking benign traffic or randomly mutating instead of exploiting vulnerabilities of ML/DL models with adversarial machine learning techniques. As a result, such attacks are very costly and inefficient, but attackers always consider attacking overhead in practice.
	Thirdly, the only traffic-space attack leveraging adversarial ML is a white-box attack\cite{hashemi2019towards}. Specifically, it assumes the attacker has all knowledge of the targeted NIDS, including used features and the ML model's architecture and parameters.
	However, such information is usually inaccessible for attackers in real-world scenarios.
	
	In summary, existing adversarial attacks (whether feature-space or traffic-space) on ML-based NIDSs failed to perform in practice due to their impractical assumptions and over-simplified traffic mutation. To bridge this gap, we face three aspects of challenges to conduct a practical study on the adversarial robustness of ML-based NIDSs:
	\begin{itemize}[]
		\item \textbf{Practicability}. How to perform a \emph{functionality-preserving} \emph{traffic-space} (under \emph{irreversible feature extraction}) attack  with extremely \emph{limited knowledge} and \emph{affordable overhead}? 
		\item \textbf{Generality}. How to propose a generic framework effective for evaluating the robustness of NIDSs using various  \emph{ML models} and \emph{features}? 
		\item \textbf{Explainability}. How to interpret the fragility and then improve the robustness of ML-based NIDSs against such adversarial attacks?
	\end{itemize}
	
	In this paper, we propose a novel practical adversarial attack by formulating it as a bi-level optimization problem with a two-step solution, which firstly solves the so-called \emph{adversarial features} with small potential traffic-space mutation overhead and secondly searches the best traffic variant with the closest feature-space distance to the adversarial features. (This addressed \emph{affordable overhead} challenge).
	
	In the first step, we extend the prior ideas of using Generative Adversarial Network (GAN) to treat the targeted ML model as a black box and search adversarial features located at the low-confidence region of the surrogate model. Based on the transferable property \cite{liu2016delving} of adversarial attacks, the attack effect can be transferable to other ML models (addressing \emph{limited knowledge} and \emph{model generality} challenges).
	In the second step, we propose a heuristic packet crafting framework to automatically mutate malicious traffic. Based on domain knowledge of feature extraction in ML-based NIDSs, we present several traffic mutation operators which can influence all summarized features without breaking malicious functionality (addressing \emph{functionality-preserving} and \emph{feature generality}).
	
	To address \emph{irreversible feature extraction} challenge, we propose an invertible abstract traffic representation called \emph{meta-info vector}.
	To address the \emph{explainability} challenge, we explain the fragility of ML-based NIDSs by quantifying the extent to which each feature is manipulated. 
	Based on this, we propose a defense scheme to improve system robustness against such attacks by removing most vulnerable features.
	
	\noindent \textbf{Contributions.} Our major contributions involve presenting a novel adversarial attack framework and extensively evaluate the robustness of ML-based NIDSs, followed by interpretation and improvement of robustness against such attacks.
	Specifically, they are elaborated as follows:
	\begin{itemize}
		\setlength{\itemsep}{0pt}
		\setlength{\parsep}{0pt}
		\setlength{\parskip}{0pt}
		\item We present the first practical traffic-space adversarial attack on ML-based NIDSs under gray and black box assumptions. 
		Compared with prior adversarial attacks, our threat model provides a more practical assumption of the attacker's ability and considers the attacking overhead.
		Compared with traditional means of crafting test-time evasive traffic, our attacks are not simply mimicking normal traffic but leverage adversarial machine learning techniques to search features located at the decision boundary of ML models and can automatically search evasive traffic mutations and are adaptive for dealing with different malicious traffic.
		
		\item We propose an explanation method for the fragile robustness of  ML-based NIDSs. Based on this, a defense scheme was proposed to improve system robustness against the proposed adversarial attacks. 
		
		\item We extensively evaluate our attack and defense method on a state-of-the-art NIDS \texttt{Kitsune}, as well as various ML-based NIDSs including six typical ML classifiers (including statistical models and deep neural networks) and two kinds of feature sets (packet-level and flow-level).
		Significant insights behind the attack are also explored through analysis.
	\end{itemize}
	
	The rest of the paper is organized as follows: 
	We summarize the related work in Section \ref{sec11} and provide backgrounds and motivations in Section \ref{sec2}.
	Section \ref{sec3} introduces the threat model and problem statement, as well as motivation and overview of our attack framework. Section \ref{sec4} and Section \ref{sec5} elaborate two steps in our attack.
	The defense schemes are provided in Section \ref{sec6}. 
	Experimental results and findings are shown in Section \ref{sec7}.
	We make discussions on limitations and improvements in Section \ref{sec8}.
	Section \ref{sec10} concludes this study.

	\begin{table}[]
		\centering
		\captionof{table}{Comparison of assumptions in adversarial attacks against ML-based NIDSs}
		
		\label{tab8}
		\resizebox{\linewidth}{!}{
			\begin{tabular}{@{}l@{ }|@{ }c@{ }c@{ }c@{ }c@{ }|c@{ }c@{}}
				\hline
				\small{\textbf{Impractical Assumptions}} &  \small{FWA} & \small{FGA} & \small{FBA} & \small{TWA} & \small{\textbf{PGA}} &\small{\textbf{PBA}$^\dag$} \\
				\hline
				\footnotesize{Unlimited traffic modification} &  
				\fullcirc &  
				\fullcirc & 
				\fullcirc & 
				\fullcirc & 
				\emptycirc & 
				\emptycirc \\
				
				\footnotesize{Directly modify features} & 
				\fullcirc &  
				\fullcirc & 
				\fullcirc & 
				\emptycirc & 
				\emptycirc & 
				\emptycirc \\
				
				\footnotesize{Knowledge on classifiers$^*$} & 
				\fullcirc &  
				\halfcirc & 
				\emptycirc & 
				\halfcirc & 
				\emptycirc & 
				\emptycirc \\
				
				\footnotesize{Knowledge on feature extractors} & 
				\fullcirc &  
				\fullcirc & 
				\fullcirc & 
				\fullcirc & 
				\fullcirc & 
				\emptycirc \\
				\hline
			\end{tabular}
		}
		\vspace{0.15em}
		
		\justifying
		\setstretch{0.85}
		\noindent \small{$\dag$}\scriptsize{FWA/FGA/FBA/TWA}: \emph{\scriptsize{\textbf{F}eature/\textbf{T}raffic-space \textbf{W}hite/\textbf{G}ray/\textbf{B}lack box \textbf{A}ttack}}
		
		\noindent \; \scriptsize{\textbf{PGA}/\textbf{PBA}} (\textbf{Ours}): \emph{\scriptsize{\textbf{P}ractical (traffic-space) \textbf{G}ray/\textbf{B}lack box \textbf{A}ttack}}
		
		\vspace{0.2em}
		\noindent $*$\emph{\scriptsize{Under this assumption, \fullcirc \ means the attacker has full knowledge about the ML model (including parameters, outputs, etc.), \halfcirc \ can only acquire the output probabilities, and \emptycirc \ neither has any knowledge nor can access to the ML model.}}
		\vspace{-2em}
	\end{table}	
	
	\section{Related Work}
	\label{sec11}
	
	We primarily introduce related work from the perspective of adversarial attacks since the robustness evaluation and primary contribution of this work is (by) proposing practical attacks.
	
	\noindent \textbf{Evasion attacks on NIDSs}.
	Evasion attacks on (N)IDS itself have been extensively studied\cite{ptacek_insertion_1998,corona_adversarial_2013,chaboya_network_2006}.
	There have been extensive studies on evading signature-based systems \cite{handley_network_2001,cheng_evasion_2012,vigna_testing_2004,mutz_experience_2003}
	and traditional anomaly-based systems
	\cite{tan_undermining_2002,kayacik_mimicry_2008,kayacik_generating_2009,fogla_polymorphic_2006,fogla_evading_2006}.
	However, such evasion attacks are only useful for signature-based NIDSs without learning models.
	
	\noindent \textbf{Adversarial attacks on (other) ML-based systems}. 
	There have been several works on adversarial attacks against ML-based systems in other domains. 
	Adversarial example in the domain of computer vision has been widely studied \cite{goodfellow_explaining_2015,moosavi-dezfooli_deepfool:_2016,carlini_towards_2017,yuan_adversarial_2019}.
	Gradient Descent method \cite{laskov_practical_2014} and Genetic Programming (GP) \cite{xu_automatically_2016} were used for evading PDF malware classifier. 
	GAN-based methods were used to attack ML-based malware detection \cite{hu_generating_2017} and real-time video classification \cite{li_stealthy_2019}. PSO-based methods were used to attack speech recognition \cite{chen2019real}, text classification \cite{zang2020word} and object detection \cite{wang2020adversarial}.
	\cite{li_textbugger:_2019} leveraged stochastic optimization method to evade text sentiment analysis. 
	However, due to the specificity of network traffic and NIDS, these methods cannot be directly applied.
	
	\noindent \textbf{Adversarial attacks on ML-based NIDSs.} We propose a terminology of Adversarial attacks on ML-based NIDSs, and list impractical assumptions of existing attacks in Table \ref{tab8}. Their specific definitions and related work are as follows:
	
	\begin{enumerate}[leftmargin=0.8em,labelsep=0.1em]
		\item \textbf{Feature-space attacks}. Feature-space attacks on evading ML-based NIDSs assume that attackers can directly modify the feature vectors. According to the attacker's knowledge of the targeted NIDS, existing feature-space attacks can be divided into three categories:
		
		\begin{itemize}[leftmargin=0.5em,labelsep=0.25em]
			\item \textbf{Feature-space White-box Attack (FWA)}. FWA requires full knowledge of the targeted NIDS.  In \cite{wang_deep_2018}, four gradient-based adversarial example attacks were directly used to evade an MLP classifier. Likewise, adversarial examples were leveraged in \cite{clements_rallying_2019} to evade \texttt{Kitsune}. Similar gradient-based methods were also used in \cite{ibitoye2019analyzing} to attack NIDSs for IoT networks, and GAN-based NIDSs \cite{piplai2020nattack}.
			
			\item \textbf{Feature-space Gray-box Attack (FGA)}. FGA requires the feedbacks of targeted classifier (without other knowledge of the classifier compared with FWA). In \cite{lin_idsgan:_2018}, a GAN-based architecture IDS-GAN was proposed to generate evasive features. In \cite{peng2019adversarial}, a boundary-based method was proposed to evade DoS  detection systems by perturbing features.
			
			\item \textbf{Feature-space Black-box Attack (FBA)}. FBA neither requires feedbacks nor any knowledges on the classifier. In \cite{apruzzese2019evaluating}, features were randomly modified to evade botnet detectors.
		\end{itemize}
		
		However, feature-space attacks are impractical since feature extraction in ML-based NIDSs is always irreversible. 
		
		\item \textbf{Traffic-space attacks}. Existing attacks that directly change network traffic can be divided into three categories:
		\begin{itemize}[leftmargin=0.5em,labelsep=0.25em]
			\item \textbf{Traffic-space White-box Attack (TWA)}. In \cite{hashemi2019towards}, a white-box attack using similar mutation operators as ours was proposed. However, their assumption that the attacker has full knowledge of the NIDS is hard to achieve in practice.
			
			\item \textbf{Random mutations}. Several mutations were proposed in \cite{stinson_towards_2008} to evade botnet detectors. Random obfuscations on traffic were proposed in \cite{homoliak_improving_2018}. However, these methods are purely stochastic and lack of theoretical guidance.
			
			\item \textbf{Traffic obfuscation}. Several traffic obfuscation methods such as \cite{wright2009traffic} have been proposed for evading traffic analysis based detection. However, they were simply mimicking benign traffic and randomly mutation without exploiting vulnerabilities of ML/DL models. 
			leveraging adversarial machine learning techniques.
		\end{itemize}
		
	\end{enumerate}

	\section{Background and Motivation}
	\label{sec2}
	
	\subsection{Targeted ML-based NIDSs}
	\label{sec2.1}
\begin{figure}[!t]
		\centering
		\includegraphics[width=0.925\linewidth]{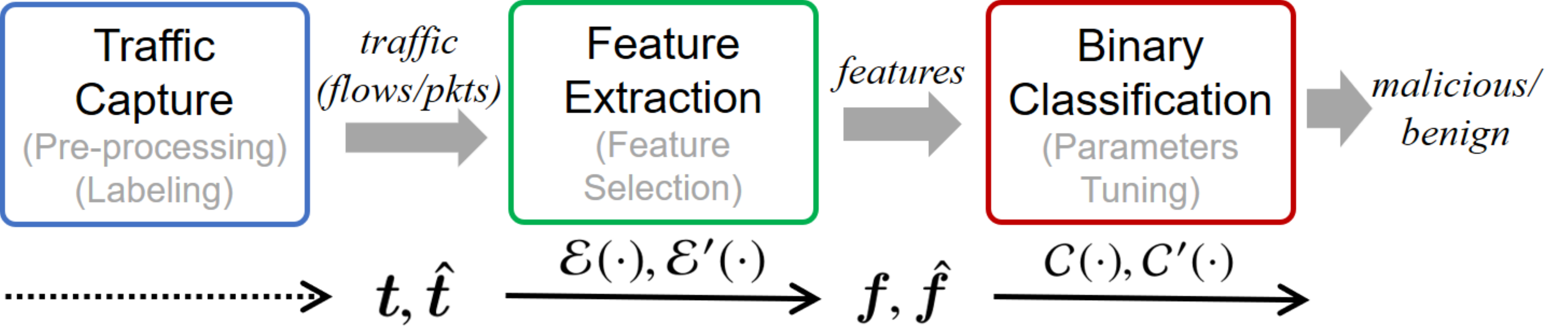}
		\caption{\small The general pipeline of ML-based NIDSs.}
		\label{fig1}
	\end{figure}

	In this study, we focused on evaluating the robustness of \emph{ML-based NIDSs} using \emph{non-payload-based} features. This kind of targeted systems is introduced as follows:

	\noindent \textbf{ML-based NIDSs}. In general, an ML-based NIDS consists of traffic capture, feature engineering, and classification as shown in Fig. \ref{fig1}. First, traffic is captured from the monitoring network. Then the feature set is extracted from raw traffic, selected, and finally fed into the ML classifier for training or detection. Note that, ML-based NIDSs need data pre-processing and manual feature extraction from raw traffic first, instead of relying solely on deep learning models for automatic feature learning. This is because network traffic is more sophisticated for having both textual and temporal characteristics, and its unstructured free text is difficult to be directly represented as variables.
	
	\noindent \textbf{Non-payload-based features}. In this study, we focus on evading ML-based NIDSs in which packets' payload is not inspected (called \emph{non-payload-based}). We think this is reasonable due to two considerations: Firstly, we find that most ML-based NIDSs are likely to use \emph{non-payload-based} features as inspecting payload is heavy and even impossible for encrypted traffic nowadays. Secondly, evading payload-based anomaly-based NIDSs has been well studied \cite{fogla_evading_2006,fogla_polymorphic_2006}.
	Note that non-payload-based NIDSs prefer to detect attacks that rely on volume and/or iteration such as DoS/DDoS (Distributed Denial of Service), scanning, brute force, and Bot/Botnet. Other attacks related to specific content such as SQL injection are out of the scope of such NIDSs. 
	
	\subsection{Adversarial Robustness Evaluation}
	\noindent \textbf{Adversarial Attacks}. Adversarial attacks on ML-based systems 
	look for constrained perturbations to normal samples, forcing the targeted model to misclassify the perturbed samples.
	In this study, we focus on one of the adversarial attacks called \emph{evasion attacks}, which can force the targeted NIDS to misclassify malicious traffic variants as benign without loss of functionality of original malicious traffic.
	
	\noindent \textbf{Robustness Evaluation}.
		In general, there are two approaches for evaluating the robustness of ML-based systems \cite{carlini_towards_2017}.
		The first one is to theoretically prove the lower bound of the robustness through formal and mathematical methods. Although lower bound is more rigorous for security analysis,
		such methods are computationally intractable to verify practical systems \cite{carlini_towards_2017}.
		By contrast, the second one, which constructs a reasonable adversarial attack and evaluates it on the targeted systems to demonstrate the upper bound of robustness (i.e., the robustness is no greater than whatever the attack can do if it succeeds), is widely used for practical systems.
	
	\subsection{Formulation and Limitations of Existing Attacks}
	\label{sec3.0}
	
	For illustration purposes, we use the function $\mathcal{E}(\cdot)$ to represent the extraction from a series of related traffic to feature vectors and $\mathcal{C}(\cdot)$ to represent ML classifiers that take feature vectors as input and output the malicious probabilities. At any certain time, we denote by $\boldsymbol{t}$ and $\hat{\boldsymbol{t}}$ two series of related \emph{original} and \emph{mutated} malicious traffic used to extract two feature vectors $\boldsymbol{f}$ and $\hat{\boldsymbol{f}}$, respectively (i.e., $\mathcal{E}(\boldsymbol{t})=\boldsymbol{f}$ and $\mathcal{E}(\hat{\boldsymbol{t}})=\hat{\boldsymbol{f}}$).
	
	As shown in Table \ref{tab8}, we argue that existing attacks are under the following impractical assumptions:
	\begin{enumerate}[leftmargin=1.5em]
		\item \textbf{Feature-space Mutation} (in FWA/FGA/FBA).
		Many previous studies merely find evasive features as solving
		$\texttt{argmin}_{\hat{\boldsymbol{f}}} \ \mathcal{C}(\hat{\boldsymbol{f}})$,
		which directly modify features' value without considering how to mutate traffic. However, features extraction is irreversible in ML-based NIDSs.
		
		\item \textbf{White-Box Knowledge} (in TWA/FWA/FGA).
		Some work assumes that classifiers' specific output is attainable, then evasion attacks can be regarded as solving an optimization problem: $\texttt{argmin}_{\hat{\boldsymbol{t}}} \ \mathcal{C}\big(\mathcal{E}(\hat{\boldsymbol{t}})\big)$. 
		However, NIDSs are always inaccessible for attackers in practice.
		
		\item \textbf{Unlimited Mutation Overhead}. Existing attacks does not limit the attacker's overhead or ability to modify traffic (e.g., with respect to traffic volume or time delay). However, attackers always consider attack overhead in practice and the traffic is unlikely to be arbitrarily modified.		
	\end{enumerate}

	In light of this, we propose a more practical adversarial attack by relaxing feature-space and white-box assumption as well as adding the constraint of overhead.
	
	\begin{table}[]
		\setstretch{0.9}
		\caption{Notations in Problem Formulation.}	
		\label{tab9}
		\footnotesize
		\centering
		\begin{tabular}{@{}c@{}|@{ }l@{}}
			\hline
			\textbf{Notation} & \multicolumn{1}{c}{\textbf{Description}} \\ \hline
			$\boldsymbol{t},\hat{\boldsymbol{t}}$  & original and mutated malicious traffic \\ \hline
			$\boldsymbol{f},\hat{\boldsymbol{f}}$  & feature vector extracted from original and mutated traffic \\ \hline
			$\mathcal{E}(\cdot),\mathcal{E}'(\cdot)$ & targeted and surrogate feature extractor \\ \hline
			$\mathcal{C}(\cdot),\mathcal{C}'(\cdot)$ & targeted and surrogate ML classifier \\ \hline
			$\mathcal{L}(\cdot,\cdot)$ & distance metric between two feature vectors \\ \hline
		\end{tabular}
	\end{table}

	\section{Attack Methodology}
	\label{sec3}
	
	In this section, we first define the threat model in a more practical setting. 
	Then, we provide the problem formulation of our attacks, followed by introduce several challenges and solutions to the problem. Finally, the overview of attack method is introduced. Some notations are listed in Table \ref{tab9}.
	
	\begin{figure*}
		\centering
		\begin{subfigure}[b]{.275\textwidth}
			\includegraphics[width=\textwidth,height=1.6cm]{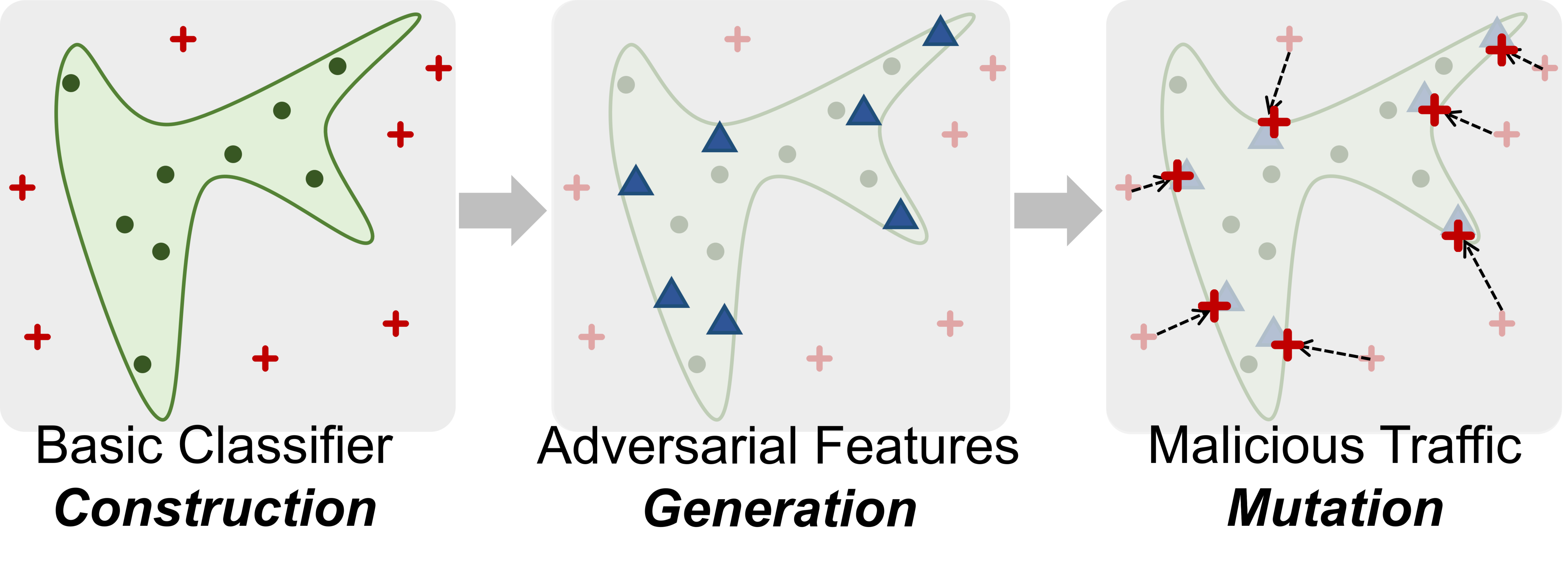}
			\vspace{-2.75ex}
			\tiny\caption{Intuitive attack framework in feature space}
			\label{fig2a}
			\vspace{3.5ex}
			
			\includegraphics[width=\textwidth,height=1.5cm]{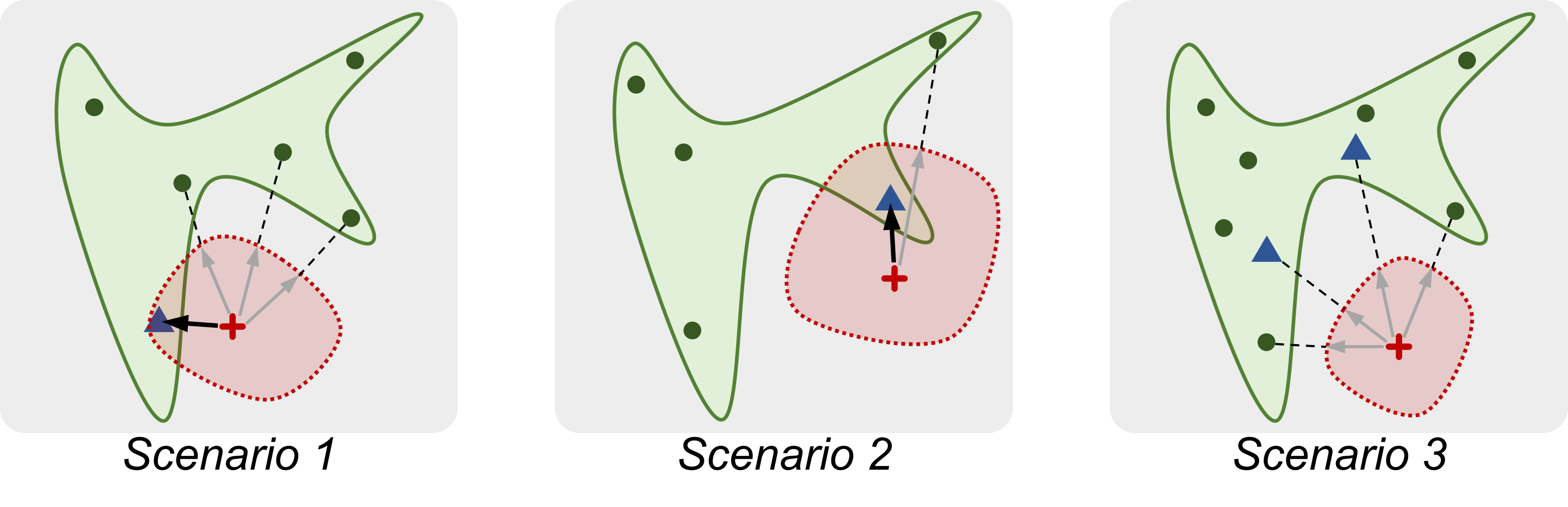}
			\vspace{-4.5ex}
			
			\caption{\footnotesize{Motivation examples of adversarial features generation}}
			\label{fig2b}
		\end{subfigure}
		\hspace{4mm}
		\begin{subfigure}[b]{.68\textwidth}
			\includegraphics[width=\textwidth,height=4.6cm]{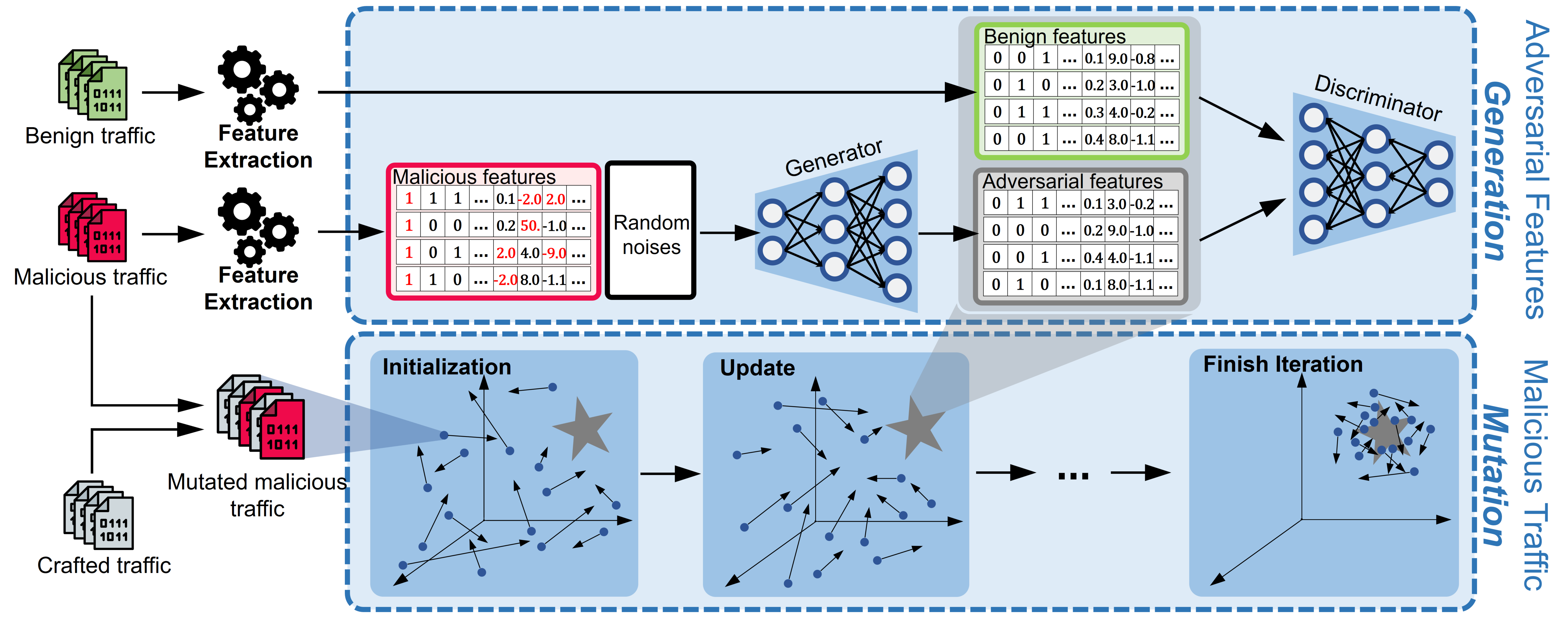}
			\vspace{-3.5ex}
			\caption{The complete framework of our evasion attack. }
			\label{fig2c}
		\end{subfigure}\qquad
		\caption{\small Attack methodology. In (\subref{fig2a}) and (\hyperref[fig2b]{b}), each plot depicts a high-dimensional feature space, in which the distribution of benign features in the targeted classifier is enclosed by a solid line with green; benign, malicious, and adversarial features are represented by small solid circles, crosses, and triangles respectively. In (\hyperref[fig2b]{b}), the limited ability/overhead of an attacker is represented by a red neighborhood.}
	\end{figure*}

	\subsection{Threat Model}
	\label{sec3.1}
	
	We consider an attacker starts with a series of traffic with malicious intent and wants to evade a ML-based NIDS using non-payload-based features (i.e., \emph{evasion attacks}).
	Unlike previous white/gray-box attacks, the attacker neither requires any knowledge about the target classifier nor its output label or probability. Unlike previous feature-space attacks, the attacker can only mutate original traffic generated from the devices he/she controls (i.e. \emph{traffic-space} attack) at an affordable \emph{overhead}. Additionally, based on the different attackers' knowledge of features in the targeted NIDS, two attacks can be performed:
	\begin{itemize}
		\setlength{\itemsep}{0pt}
		\setlength{\parsep}{0pt}
		\setlength{\parskip}{0pt}
		\item \textbf{Practical Gray-box Attack (PGA)}. This assumes features used in the targeted NIDS are known by the attacker. In other worlds, the attacker can build the same feature extractor as the targeted NIDS and use it to extract features exactly. This may seem extreme, but the features used in ML-based NIDSs are often published \cite{umer2017flow,davis_data_2011,mirsky_kitsune:_2018,draper-gil_characterization_2016,tavallaee_detailed_2009}.
		\item \textbf{Practical Black-box Attack (PBA)}. We assume a more practical case, in which the attacker has very limited or even no knowledge about the features used in the targeted NIDS. In this case, the attacker can only build a surrogate extractor based on the domain knowledge.
	\end{itemize}
	
	\subsection{Practical Traffic-space Evasion Attack Problem}
	
	According to the threat model, firstly, we relax two assumptions in Section \ref{sec3.0} by training a substitute classifier $\mathcal{C}'(\cdot)$ with probabilistic output to approximate $\mathcal{C}(\cdot)$. This also solves the problem that some ML models without continuous output values (such as Isolation Forest) are difficult to optimize.
	Secondly, we build the surrogate feature extractor $\mathcal{E}'(\cdot)$. 
	As for PGA, $\mathcal{E}'(\cdot)$ is exactly the same as $\mathcal{E}(\cdot)$, while is simulated as for PBA. How to build $\mathcal{C}'(\cdot)$ and $\mathcal{E}'(\cdot)$ will be introduced later.
	Additionally, we denote the mutation operation as $\mathcal{M}(\cdot)$ which can transform original traffic $\boldsymbol{t}$ to a set consisting of all possible mutated traffic $\hat{\boldsymbol{t}}$. We say a mutation operation is \emph{safe} (denoted by $\mathcal{M}_{s}(\cdot)$) if the mutation can preserve the malicious functionality of $\boldsymbol{t}$. Besides, we also consider mutation overhead budget 
	through an additional term to minimize the distance (denoted by $\mathcal{L}^\ast(\cdot,\cdot)$ with a weight coefficient $\lambda$) between $\hat{\boldsymbol{t}}$ and $\boldsymbol{t}$. Therefore, the attack problem can be formulated as solving:
	\vspace{-0.35em}
	\begin{equation}
	{\operatorname{argmin}_{\hat{\boldsymbol{t}}} \ \mathcal{C}'\big(\mathcal{E}'(\hat{\boldsymbol{t}})\big) + 
		\lambda \mathcal{L}^\ast(\hat{\boldsymbol{t}},\boldsymbol{t})
		\quad {\rm s.t.} 
		\ \ \hat{\boldsymbol{t}} \in \mathcal{M}_{s}(\boldsymbol{t}).
		\label{eq1}}
	\vspace{-0.35em}
	\end{equation}
	
	\vspace{-.5em}
	\subsection{Challenges and Solutions}	
	Problem (\ref{eq1}) is hard to solve due to several challenges, such as the highly non-linear constraint, non-differentiable $\mathcal{E}'(\cdot)$, and hard-to-define $\mathcal{L}^\ast(\cdot,\cdot)$.
	Next, we will introduce how we transform the problem and solve it. 
	
	\subsubsection{\textbf{Bridging the gap of traffic-feature space}}
	As mentioned before, we observe that the attacker's so-called \emph{\textbf{overhead budget}} or \emph{\textbf{ability}} to mutate traffic is always limited in practice.
	For one thing, an attacker is likely to have a \emph{budget} of \emph{overhead} (such as the extra time and crafted traffic volume to evade detection).
	For another, the attacker's \emph{ability} to modify traffic is also limited. For examples, excessively increasing the interval time will cause the connection timeout, and injecting excessive traffic will occupy a lot of bandwidth and may be perceived by the victim.
	Considering the mutation overhead, directly searching the qualified $\hat{\boldsymbol{t}}$ from the sophisticated traffic space is inefficient due to the gap of traffic-feature space. 

	To address this problem, we introduce so-called \textbf{\emph{adversarial feature}} denoted with $\boldsymbol{f}^{\star}$ to bridge this gap. 
	We transform problem (\ref{eq1}) into the following bi-level optimization problem:
	\begin{align}
	\operatorname{argmin}_{\hat{\boldsymbol{t}}} \ & \ \mathcal{L}\Big( \mathcal{E}'\big(\hat{\boldsymbol{t}}\big),\boldsymbol{f}^{\star} \Big) \label{eq2} \\ 
	{\rm s.t.} \ & \  \boldsymbol{f}^{\star} \ = \  \operatorname{argmin}_{\boldsymbol{f}^{\star}} \  
	\mathcal{L}\Big( \boldsymbol{f}^{\star},\mathcal{E}'\big(\boldsymbol{t}\big) \Big) \label{eq3} \\ 
	\ & \  \mathcal{C'}\big(\boldsymbol{f}^{\star}\big)<h \label{eq6} \\
	\ & \ \hat{\boldsymbol{t}} \in \mathcal{M}_{s}(\boldsymbol{t}),	\label{eq7}
	\end{align}
	where $\mathcal{L}(\cdot,\cdot)$ is a distance metric between two feature vectors and $h$ is the anomaly threshold for classification. Equation (\ref{eq6}) means that $\boldsymbol{f}^{\star}$ is classified as benign.
	The high-level idea to solve the above problem is to separately solve the lower and upper-level objective function: We firstly solve adversarial feature $\boldsymbol{f}^{\star}$ in (\ref{eq3}) under the constraint (\ref{eq6}) in feature space, and then use the solved $\boldsymbol{f}^{\star}$ to search  $\hat{\boldsymbol{t}}$ in (\ref{eq2}) under the constraint (\ref{eq7}). 
	To give an intuition, Fig. \ref{fig2a} depicts these two steps from the perspective of feature space.
	Firstly, for each malicious feature, an $\boldsymbol{f}^{\star}$ is produced which can be not only classified as benign but also as close as possible to the malicious feature. 
	Secondly, original malicious traffic is mutated to transfer its features to the closest adversarial(/benign) ones.
	
	In general, $\boldsymbol{f}^{\star}$ lies on the low-confidence region of the classifier. Note that, $\boldsymbol{f}^{\star}$ is in feature space and does not need to correspond to traffic space.
	Our key observation is that \emph {mutation overhead in traffic-space is correlated to the distance in feature-space} (Details and proof are in Appendix \ref{corr}). Thus, $\boldsymbol{f}^{\star}$ in feature space can reduce the overhead of mutating traffic. Fig. \hyperref[fig2b]{2b} also provides examples to demonstrate the necessity of $\boldsymbol{f}^{\star}$.
	Without guidance of adversarial features, a malicious feature may eventually fail to reach the nearest benign space (Scenario 1) or miss the transient benign space (Scenario 2). 
	
	\subsubsection{\textbf{Solving adversarial features $\boldsymbol{f}^{\star}$}}
	Adversarial features generation needs to be: i) model-agnostic---we assume the attacker has no knowledge about the ML classifier; and ii) efficient---there may be tons of malicious/traffic in practice.
	Inspired by previous adversarial attacks\cite{hu_generating_2017,lin_idsgan:_2018}, GAN \cite{goodfellow_generative_2014} is highly competent to generate adversarial features since i) the \emph{discriminator} can be trained as a substitute for the targeted classifier (i.e., $\mathcal{C}'(\cdot)$ in (\ref{eq1}) and (\ref{eq6})), thus can conduct model-agnostic attack; ii) once the \emph{generator} is trained, it can generate $\boldsymbol{f}^{\star}$ efficiently for any malicious feature.
	Details will be introduced in Section \ref{sec4}. 
	
	\subsubsection{\textbf{Solving mutated traffic $\hat{\boldsymbol{t}}$}}
	Network traffic is unstructured free text thus cannot directly participate in the numerical calculation, so we vectorize traffic as high dimensional so-called \emph{meta-info} vectors denoted with $\boldsymbol{x}$ involving meta information of packets' header. We propose several \emph{safe} mutation operations on $\boldsymbol{x}$ to release the constrain (\ref{eq7}).
	Therefore, we get the final unconstrained optimization problem:
	\begin{align}
	\operatorname{argmin}_{\boldsymbol{x}} \ & \ \mathcal{L}\Big( \mathcal{E}'\big(\mathcal{R}(\boldsymbol{x})\big),\boldsymbol{f}^{\star} \Big), \label{eq10}
	\end{align}
	where $\mathcal{R}(\cdot)$ represents rebuild traffic from meta-info vectors, i.e., $\mathcal{R}(\boldsymbol{x})=\hat{\boldsymbol{t}}$. However, most of dimensions in meta-info vectors are discrete and $\mathcal{E}'(\cdot)$ is non-differentiable, thus problem (\ref{eq2}) is indeed a hard combinatorial optimization task (NP-complete). Hence, we resort to swarm intelligence algorithms to find approximate solutions. We employ PSO \cite{kennedy_particle_2010} due to its great adaptability on dealing with high dimensional and discrete tasks\cite{chen2019real, zang2020word,wang2020adversarial}. Details will be introduced in Section \ref{sec5}. 

	\subsection{Overview of Attack Method}
	\label{sec3.3}
	
	Now we briefly introduce the overview of our attack methodology from a more operational way. The workflow is illustrated in Fig. \ref{fig2c}, including the following two steps, corresponding to the aforementioned bi-level optimization:
	\begin{itemize}
		\item \textbf{Adversarial Features Generation}: We assume an attacker wants to launch some activities, which will induce a series of \emph{malicious} traffic. First, the attacker needs to collect some \emph{benign} traffic in the network he/she controls. Then, two kinds of traffic are extracted into features by the surrogate extractor, and fed into our GAN model. After the training phase, the generator is capable to generate adversarial features.
		
		\item \textbf{Malicious Traffic Mutation}: After generating adversarial features, we employ PSO with predefined \emph{safe} operators to mutate malicious traffic automatically. Each particle in the swarm represents a vector consists of meta-info of mutated traffic.
		The swarm is iteratively searching the traffic-space under the guidance of the temporary best particle whose features are most similar to the adversarial feature. Finally, the best particle is selected after several iterations.
	\end{itemize}
	
	The details of the above two steps are elaborated in next two sections \ref{sec4} and \ref{sec5}, respectively.

	\section{Generating Adversarial Features}
	\label{sec4}
	We now introduce the procedure of generating adversarial features. Our enhanced GAN model is shown in Fig. \ref{fig2c} on the top, which consists of a generator and a discriminator.
	
	\noindent \textbf{Generator.} The generator is a feed-forward neural network whose aim is to transform a malicious feature into its adversarial version. It takes the concatenation of a malicious feature vector $\boldsymbol{f}$ and a noise vector $\boldsymbol{z}$ from Gaussian distribution $p_{\boldsymbol{z}}(\boldsymbol{z})$ as input and outputs a generated feature vector represented by $G(\boldsymbol{f}, \boldsymbol{z})$. To train the generator, its loss function is defined as:
	\begin{equation}
	l_{G}=\mathbb{E}_{\boldsymbol{f} \in \mathbf{F}_{mal}, \boldsymbol{z} \sim p_{\boldsymbol{z}}(\boldsymbol{z})}[\log D(G(\boldsymbol{f}, \boldsymbol{z}))+\mathcal{L}(\boldsymbol{f}, G(\boldsymbol{f}, \boldsymbol{z}))],
	\label{eq4}
	\end{equation}
	where $\mathbf{F}_{mal}$ is the set of original malicious features. $l_{G}$ should be minimized with respect to the weights in the generator's network. In this study, we extend prior GANs by additionally computing a construct error $\mathcal{L}(\cdot,\cdot)$ between the input and output. In this study, we use the root mean square error
	$\small{\operatorname{RMSE}\left(\boldsymbol{x}, \boldsymbol{x}^{\prime}\right) = \sqrt{\sum_{i=1}^{n_{d}}\left(\boldsymbol{x}_{i}-\boldsymbol{x}_{i}^{\prime}\right)^{2} / n}}$ where $n_{d}$ is the dimensionality of the input and output features.
	Thus, the generated features can mimic the distribution of benign features while approaching malicious ones.
	
	\noindent \textbf{Discriminator.} The discriminator trained as the surrogate classifier (i.e., $\mathcal{C}'(\cdot)$) is used to distinguish generated features from benign ones. It is also a feed-forward neural network whose input consists of benign feature vectors and generated malicious feature vectors from the generator, and its output is a probability of determining an input vector is generated (also malicious). The discriminator is trained to maximize the output of generated input vector while minimize the output of benign input vectors. Thus, its loss function is:
	\begin{equation}l_{D}=-\mathbb{E}_{\boldsymbol{f} \in \mathbf{F}_{ben}} \log (1-D(\boldsymbol{f}))-\mathbb{E}_{\boldsymbol{f} \in \mathbf{F}_{gen}} \log D(\boldsymbol{f}),
	\end{equation}
	where $\mathbf{F}_{ben}$ is the set of features extracted from benign traffic collected in the network the attacker controls beforehand. $\mathbf{F}_{gen}$ is the set of features generated by the generator.
	
	The training process is an iterative and mutual optimization between the generator and discriminator until a convergence. Then, the generated features from the generator can work as adversarial features $\mathbf{F}_{adver}$.

	\section{Mutating Malicious Traffic}
	\label{sec5}
	In this section, we introduce how to automatically mutate traffic through our PSO-based method. First, we design the mutation operators on malicious traffic. Then, we introduce the vectorization from traffic to meta-info vectors. Finally, we propose the PSO-based traffic mutation algorithm.
	
	\subsection{Basic Traffic Mutation Operators}
	\label{sec5.2}
	To solve the evasive traffic mutation, we first design some basic parameterized \emph{mutation operators}.
	The mutation operators should be able to affect as many types of features as possible to be generic, and also should be
	functionality-preserving (or safe, recall $\mathcal{M}_{s}(\cdot)$ in (\ref{eq7})) and stealthy to prevent being perceived by victims. 
	However, it is challenging when attacker has limited or even no knowledge of the features used in the targeted system (i.e., PBA).
	To address this challenge, we propose a high-level summarization methodology of features used in non-payload-based NIDSs \cite{umer2017flow,davis_data_2011,mirsky_kitsune:_2018,draper-gil_characterization_2016,tavallaee_detailed_2009} by dividing features extracted from network traffic into two dimensions---Temporal and Spatial: (More details are in Appendix \ref{app1})
	\begin{itemize}
		\setlength{\itemsep}{0pt}
		\setlength{\parsep}{0pt}
		\setlength{\parskip}{1pt}
		
		\item Temporal features are related to the timing of traffic, such as the inter-arrival time of packets.
		\item Spatial features are further divided into Global and Local Spatial features:
		\begin{itemize}
			\setlength{\itemsep}{0pt}
			\setlength{\parsep}{0pt}
			\setlength{\parskip}{1pt}
			\item  Global Spatial features are the overall characteristics of traffic (for example, the volume of traffic, such as the number of bytes or the total number of packets).
			\item  Local Spatial features are related to the content of packets. Since we only focus on the packet headers and certain protocols, the types of Local Spatial features are limited.
		\end{itemize} 
	\end{itemize} 
	
	Then, we design mutation operators which can affect all summarized high-level features (Fig. \ref{figapp} in Appendix \ref{app1} provides an intuitive illustration) while preserving the functionality. Specifically, they consist of modifying original malicious traffic and injecting/adjusting crafted stealthy traffic:
	
	\vspace{0.2em}
	\noindent \textbf{Original malicious traffic modification.} We ensure that the original traffic will not be deleted and the order of packets will not be changed. Hence the only mutation operator is:
	{\small\textsf{
			\begin{enumerate}[label*=(\alph*),leftmargin=1.75em]
				\item Altering the \textbf{interarrival time} of packets in original traffic 
	\end{enumerate}}}

	\begin{table}[]
		\caption{Crafted traffic generation method}
		\label{tab11}
		\small
		\centering
		\begin{tabular}{@{}c@{}|@{}l@{}}
			\hline
			\bf{Traffic type}&\multicolumn{1}{c}{\bf{Generation methods}}
			\\ \hline
			Any & \rule{0pt}{12.5pt}\footnotesize
			\renewcommand\arraystretch{0.8}\begin{tabular}[c]{@{ }l@{}}
				Subtly assigning the \textbf{TTL} field so that the NIDS can receive \\ the crafted packet but the victim cannot\cite{hashemi2019towards}.\end{tabular}
			\\ \hline
			TCP & \rule{0pt}{12.5pt}\footnotesize
			\renewcommand\arraystretch{0.8}\begin{tabular}[c]{@{ }l@{}}
				Requesting the establishment (i.e., send \textbf{SYN}) of an \\ established or establishing connection again.\end{tabular}
			\\ \cdashline{1-2}[1.75pt/2pt]
			\multirow{2}*{\begin{tabular}[c]{@{}c@{ }}\rule{0pt}{12pt} TCP\\(established)\end{tabular}} 
			& \rule{0pt}{12.5pt}\footnotesize \renewcommand\arraystretch{0.8}\begin{tabular}[c]{@{ }l@{}}
				Packets with smaller (already acknowledged) or larger \\ \textbf{sequence number} than expected.\end{tabular}
			\\ \cdashline{2-2}[1.75pt/2pt]
			& \rule{0pt}{12.5pt}\footnotesize \renewcommand\arraystretch{0.8}\begin{tabular}[c]{@{ }l@{}}
				Packets with smaller or larger \textbf{acknowledge number} than \\ expected.\end{tabular}
			\\ \hline
			UDP/ICMP & \rule{0pt}{12.5pt}\footnotesize \renewcommand\arraystretch{0.8}\begin{tabular}[c]{@{ }l@{}}
				Padding packets' payload with semantical-free content (such \\ as randomly padding).\end{tabular}
			\\ \hline
			ICMP & \rule{0pt}{9pt}\footnotesize \renewcommand\arraystretch{0.8}\begin{tabular}[c]{@{ }l@{}}
				Packets with deprecated type or code field. \end{tabular}
			\\ \hline
			
		\end{tabular}	
	\end{table}

	\noindent \textbf{Crafted stealthy traffic injection.} It is non-trivial to determine packer headers' content of crafted traffic. Firstly, we can only craft traffic sent from the attacker, and some fields (MAC/IP/port) in crafted packets need to be consistent with that of original packets nearby; otherwise the crafted packets cannot affect features extracted from original packets. Secondly, the assignment of other fields in header must meet the following requirements: 1) it will not compromise the maliciousness of the original traffic; 2) it will not cause the protocol semantics or communication violation (such as connection breakdown of TCP traffic); 3) 
	it will not induce responses by the victim (for stealth and consistency of replay). In light of these requirements, we list optional methods for generating crafted traffic in Table \ref{tab11}. We note that a previous method used in \cite{hashemi2019towards} by modifying TTL requires the knowledge of the victim's network topology, which is extremely strict. Hence, we propose other methods for different types of traffic without additional knowledge.
	
	\noindent \textbf{Crafted traffic adjustment.}
	There are several adjustments for crafted packets after being injected:
	
	\vspace{-1.75ex}
	{\small\textsf{
			\begin{enumerate}[label*=(b\arabic*),leftmargin=2.25em]
				\item Altering the \textbf{interarrival time} of packets in crafted traffic
				\item Altering the \textbf{\# protocol layer} of packets in crafted traffic
				\item Altering the \textbf{payload size} of packets in crafted traffic
	\end{enumerate}}}

	\begin{figure}[]
		\centering
		\includegraphics[width=0.97\linewidth]{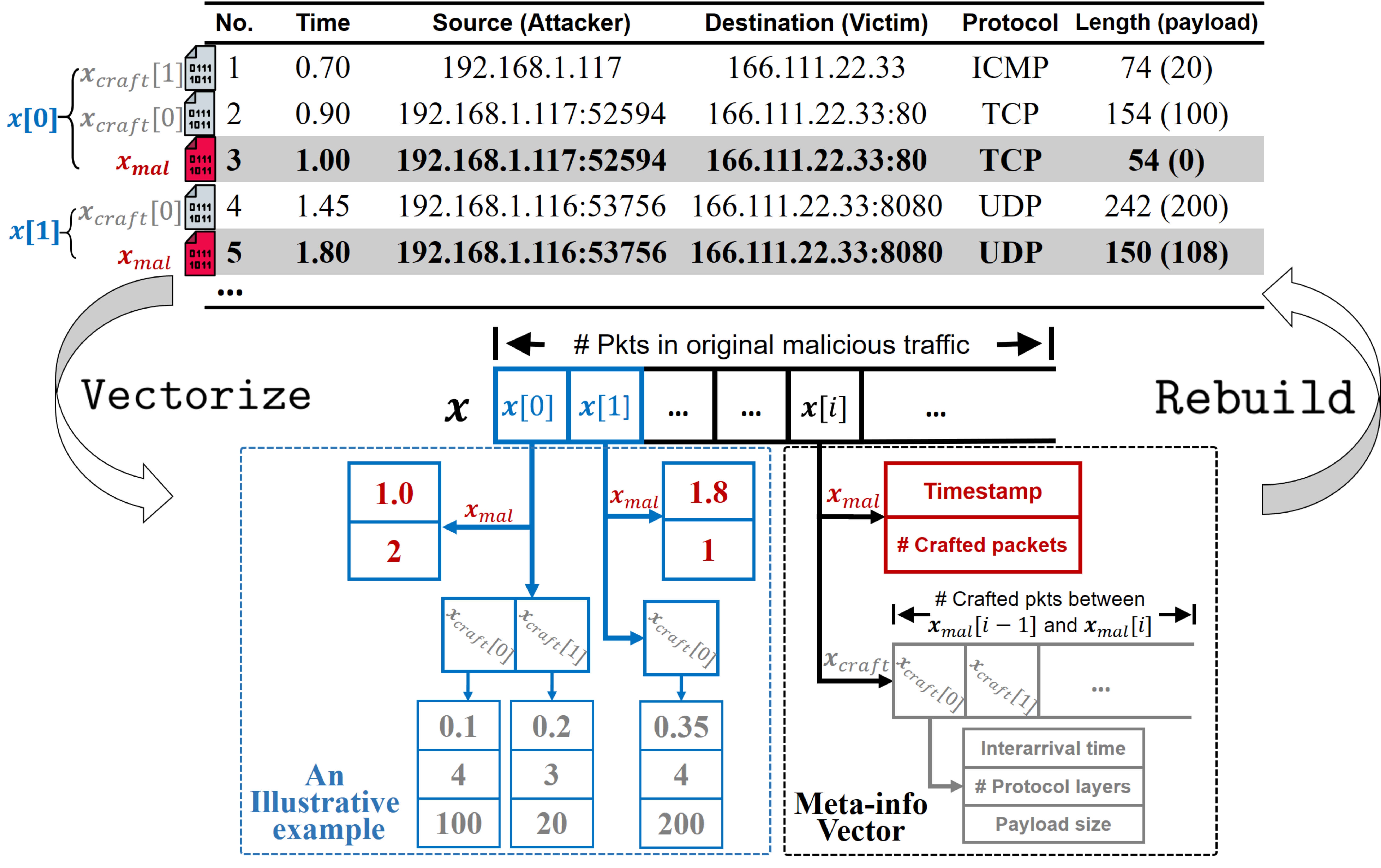}
		\caption{\small Vectorization and Rebuilding between vectors and traffic.}
		\label{fig4}
	\end{figure} 
	
	\subsection{Meta-information Vectorization}
	\label{sec5.3}
	
	To facilitate numerical operations on structured traffic data, we \emph{vectorize} traffic into \emph{meta-info vectors} containing meta-information of original traffic. Note that unlike feature extraction, this vectorization is invertible, which means that it is easy to \emph{rebuild} traffic from meta-info vectors. Meanwhile, the aforementioned mutation operators on traffic need to be reflected in the vectors. Details of the meta-info vectors and an illustrative example about vectorization and rebuilding between vectors and traffic are shown in Fig. \ref{fig4}, where $\boldsymbol{x}$ denotes the meta-info vectors.
	
	To illustrate the meaning of each dimension in $\boldsymbol{x}$ and how 
	they reflect the mutation operators, an $\boldsymbol{x}[i]$ is further divided into $\boldsymbol{x}_{mal}$ and $\boldsymbol{x}_{craft}[]$ used to represent a packet in original malicious traffic and several crafted packets right in front of it in time, respectively. The $\boldsymbol{x}_{mal}$ contains two parts: {\small\textsf{Timestamp}} corresponds to Mutation {\small\textsf{(a)}}, and {\small\textsf{Number of crafted packets}} determines the size of the list $\boldsymbol{x}_{craft}$.
	Each craft packet denoted with $\boldsymbol{x}_{craft}[i]$ contains three parts:  {\small\textsf{Interarrival time}} related to Mutation {\small\textsf{(b1)}} is the time interval from the previous packet; {\small\textsf{\# protocol layers}} corresponding to Mutation {\small\textsf{(b2)}} refers to its number of layers in the TCP/IP protocol; {\small\textsf{Payload size}} directly reflects Mutation {\small\textsf{(b3)}}.
	
	\noindent \textbf{Overhead budget}. In this study, we limit the attacker's overhead budget from two aspects. The first overhead denoted with $l_{c}$ is the rate of the number of crafted packets to that of original packets. The second overhead denoted with $l_{t}$ is the rate of time elapsed of mutated traffic to that of original traffic. In other words, the crafted packets number and time elapsed of mutated traffic must no more than $l_{c}$ and $l_{t}$, respectively.

	\noindent \textbf{Dimensionality}. Note that for the convenience of illustration, we split $\boldsymbol{x}$ into $\boldsymbol{x}_{mal}$ and $\boldsymbol{x}_{cal}$ with meaningful dimensions. For implementation, Meta-info vectors $\boldsymbol{x}$ are flattened into well-structured $D$-dimensional vectors. $D$ can be computed according to the composition of $\boldsymbol{x}$:
	\begin{equation}
	D = N_{pkts}(3l_{c}+2), \label{eq9}
	\end{equation}
	where $N_{pkts}$ is the number of packets in malicious traffic.

	\subsection{PSO-based Automatic Traffic Mutation}
	We now present our algorithm for automatically searching the best traffic mutants based on PSO (in (\ref{eq10})). The general framework of PSO is shown in Fig. \ref{fig2c}.
	Each particle represents a candidate traffic mutant, which consists of a \emph{position} and \emph{velocity} vector.
	Each position vector in the swarm is exactly the meta-info vector denoted by $\boldsymbol{x}^n$ ($n\in\{1,2,\cdots,N_{swarm}\}$ where $N_{swarm}$ is the number of particles in the swarm), and the velocity vector denoted by $\boldsymbol{v}^n$ shares the same shape with $\boldsymbol{x}^n$. Then we describe
	our PSO-based algorithm step by step. (We also provide the algorithm format in Appendix \ref{app3}.)
	
	\subsubsection{\textbf{Initialization}} At the very beginning,
	to sufficiently disperse initial particles in the search-space, fields in $\boldsymbol{x}_{craft}[]$ and {\small\textsf{\# crafted pkts}} in $\boldsymbol{x}_{mal}$ are randomly initialized within the valid range.
	As for {\small\textsf{Timestamp}} of $\boldsymbol{x}_{mal}$, we divide the maximum interarrival time (related to $l_{t}$) between every two original packets into $m$ equal parts, and {\small\textsf{Timestamp}} is randomly selected from these $m$-section points.
	And $\boldsymbol{v}^n$ is filled with $0$ initially.
	
	\subsubsection{\textbf{Effectiveness Evaluation}} In each iteration, each particle is evaluated to judge how close they are to the objective according to its position $\boldsymbol{x}^n$ ($n\in\{1,2,\cdots,N_{swarm}\}$). The effectiveness is defined as (\ref{eq11}) according to (\ref{eq10}):
	\begin{equation}
	Effectiveness(\boldsymbol{x}^n) = \mathbb{E}_{\boldsymbol{f}^{\star} \in \mathbf{F}_{adver}} \| \mathcal{E}'(\mathcal{R}(\boldsymbol{x}^n))-\boldsymbol{f}^{\star} \|, \label{eq11} 
	\end{equation}
	where can be viewed as three steps:
	First, mutated traffic is rebuilt (denoted by $\mathcal{R}$) from the position $\boldsymbol{x}^n$. Original traffic is directly retrieved after a replacement with {\small\textsf{Timestamp}} in $\boldsymbol{x}_{mal}$.
	As for crafted packets, after randomly determining their protocol type with reference to {\small\textsf{\# protocol layers}} in $\boldsymbol{x}_{craft}[i]$, they can be rebuilt through the methods listed in \ref{sec5.2}. Fig. \ref{fig4} also shows an example of traffic rebuilding.
	Second, mutated traffic is extracted (denoted by $\mathcal{E}'$) into features through the surrogate extractor. Third, the distance between extracted and adversarial features is computed as the effectiveness.
	
	\subsubsection{\textbf{Update}} After computing the effectiveness of $\boldsymbol{x}^n$, the position it has reached with the best effectiveness is recorded as its \emph{individual best} position denoted by $\boldsymbol{b}^n$, and the best position in the swarm called \emph{global best} position is denoted by $\boldsymbol{g}$.
	Then, each particle updates its velocity $\boldsymbol{v}^n$ as follows:
	\begin{equation}
	\boldsymbol{v}^n_d = \omega\boldsymbol{v}^n_d + r_{1}c_{1}(\boldsymbol{b}^n_d-\boldsymbol{x}^n_d) + r_{2}c_{2}(\boldsymbol{g}_d-\boldsymbol{x}^n_d), 
	\end{equation}
	where $\omega,c_{1},c_{2}$ are weight coefficients of three terms and $r_1,r_2$ are random coefficients, and $d\in\{1,2,\cdots,D\}$ ($D$ is the same as (\ref{eq9})).
	And then position $\boldsymbol{x}^n$ is updated with the new $\boldsymbol{v}^n_d$:
	\begin{equation}
	\boldsymbol{x}^n_d = \texttt{clip}_d(\boldsymbol{x}^n_d+\boldsymbol{v}^n_d), 
	\end{equation}
	where the function \texttt{clip} limits the dimension beyond the value range to the minimum/maximum. Besides, discrete dimensions of $\boldsymbol{x}$ (e.g., {\small\textsf{\# crafted pkts}}) are discretized by approximating them to the nearest discrete values.  
	
	\subsubsection{\textbf{Finish Iteration}}
	The above two steps (evaluation and update) perform iteratively for a fixed number of iterations denoted by $N_{iter}$. Then, traffic rebuilt from the current global best position $\boldsymbol{g}$ becomes the best evasive mutant.

	\section{Defense Scheme}
	\label{sec6}
	To defend against the proposed attacks and improve the robustness of ML-based NIDSs, we introduce two probable methods in prior work, and then propose a novel defense scheme named \emph{adversarial feature reduction}.
	
	\noindent \textbf{Adversarial training}\cite{goodfellow_explaining_2015}.
	This is a promising method widely used to defend against adversarial examples in the image domain by retraining the classifiers with correctly-labeled adversarial examples. However in our traffic-space attack, it can only reduce the attack effectiveness by limiting the generation of adversarial features.
	
	\noindent \textbf{Feature selection}\cite{guyon_introduction_2003}. This is an important step in feature engineering to remove redundant/irrelevant dimensions of features used in ML models, which can effectively improve detection performance and robustness.
	
	\noindent \textbf{Adversarial feature reduction}. We propose a novel scheme to explain and defend against such traffic-space adversarial attacks. In a nutshell, we proactively simulate the proposed attack and then calculate the degree to which the value of each feature dimension in the mutated traffic is close to the adversarial features compared to original value (see Appendix \ref{app21} for details). The proximity rates of each feature dimension can be viewed as the adversarial robustness scores. Our main claim is the high dimensionality of features gives attackers an opportunity to exploit some vulnerable dimensions to evade detection.
	Hence, we propose an intuitive defense scheme by deleting partial feature dimensions with low robustness scores.
	
	\newcommand{\tabincell}[2]{\begin{tabular}{@{}#1@{}}#2\end{tabular}}  

	\section{Experimental Evaluation} 
	\label{sec7}
	In this section, we first introduce the experimental settings\footnotemark[1] in \ref{sec7.1}.
	In \ref{sec7.2} and \ref{sec7.3}, we compare the effectiveness of our attacks with baseline attacks. Several experiments are conducted using different kinds of malicious traffic (in \ref{sec7.2}) and different NIDSs (in \ref{sec7.3}).
	We evaluate the performance of our PBA attacks in \ref{sec7.4}.
	Execution cost and impact of parameters are measured in \ref{sec7.5}.
	We verify our attack is functionality-preserve in \ref{sec7.6}.
	Finally, defense methods are evaluated in \ref{sec7.7}.
	
	\footnotetext[1]{To strengthen scientific reproducibility, we provide a supplemental material consists of detailed settings of ML/DL models in our experiments, as well as the implementation code of adversarial packet crafting and optimized implementation of targeted NIDSs, which are available at: \textbf{\url{https://github.com/dongtsi/TrafficManipulator}}}

	\subsection{Experimental Settings}
	\label{sec7.1}

	\noindent \textbf{Datasets}. Table \ref{tab2} summarizes the information of traffic sets used in this study, including six well-known attacks from two up-to-date traffic datasets.
	\emph{Kitsune Dataset}\cite{mirsky_kitsune:_2018} was used to evaluate \texttt{Kitsune} by proactively performing a number of attacks in their video surveillance network. \emph{CIC-IDS2017} \cite{sharafaldin_toward_2018} collected traffic for common attacks in a large-scale testbed, which covers all common devices and middleboxes. 
	Note that, we additionally select different benign traffic with the same packet number of test set for adversarial feature generation.

	\noindent \textbf{Targeted NIDSs}.
	Firstly, \texttt{Kitsune}\cite{mirsky_kitsune:_2018} is evaluated as the state-of-the-art off-the-shelf NIDS, which consists of a packet-based feature extractor and Autoencoder model.
	Secondly, NIDSs using different ML classifiers (including traditional statistical learning \cite{qayyum2005taxonomy} and deep neural networks) and different kinds of features are extensively evaluated (including models and extractors used in prior work on flow-level anomaly/intrusion detection \cite{umer2017flow,sperotto2010overview}):
	\begin{itemize}
		\item \textbf{Feature extractors}: We evaluate two representative feature extractors: \emph{AfterImage}\cite{mirsky_kitsune:_2018} is a \emph{packet-based} extractor in \texttt{Kitsune}. It computes incremental statistics of packet's size, count and jitter in various damped time windows. \emph{CICFlowMeter}\cite{draper-gil_characterization_2016} is a \emph{flow-based} extractor. It extracts several statistics (e.g., size, count, and duration) of connections.
		
		\item \textbf{ML classifiers}: We apply six classifiers that are widely used in related work to comprehensively cover ML models \cite{hodo_shallow_2017,gulghane2019survey,ferrag2020deep}.
		\emph{KitNET} is a deep, unsupervised, and ensemble Autoencoders used in \texttt{Kitsune}.
		\emph{Multi-Layer Perceptron} (\emph{MLP}) represents supervised deep learning models.
		\emph{Logistics Regression} (\emph{LR}), \emph{Decision Tree} (\emph{DT}) and \emph{Support Vector Machine} (\emph{SVM}) represent supervised traditional ML models.
		\emph{Isolation Forest} (\emph{IF}) represents unsupervised anomaly detection model.
	\end{itemize}
	
	\noindent \textbf{Baseline attacks}. For one thing, we find that previous attacks against signature-based NIDSs (like \cite{ptacek_insertion_1998,handley_network_2001,cheng_evasion_2012}) and 
	traditional anomaly-based NIDSs (like \cite{fogla_polymorphic_2006}) perform nearly no evasive effect. This is because these methods focused more on manipulating the payload.
	For another, feature-space attacks (e.g.,\cite{wang_deep_2018,clements_rallying_2019}) including FWA/FGA/FBA cannot participate either since they did not mutate the traffic. Hence, we existing employ traffic-space attacks as baselines:
	\begin{itemize}
		\item \textbf{Random Mutation}. Note that randomly mutating traffic is not a baseless weak attack, but appears in published works \cite{stinson_towards_2008, homoliak_improving_2018}. We use two random mutation methods: {\small\textsf{\textbf{Random-ST}}} is randomly spreading interval-time between packets; {\small\textsf{\textbf{Random-Dup}}} is randomly duplicating partial original traffic. As for other methods, \emph{packet injection} is not used since we find it has no effect on all traffic sets; \emph{deleting/reordering packets} compromises the functionality. 
		
		\item \textbf{Traffic-space White-box Attack ({\small\textsf{TWA}})}. The only work of TWA we find is \cite{hashemi2019towards}, which uses similar mutation operators as ours. Since attackers have full knowledge of the targeted NIDSs in their assumption, the output probabilities are directly used as the optimization objective. 
		
	\end{itemize}

	\begin{table}[]
		\caption {Attack traffic datasets.}
		\label{tab2}
		\footnotesize
		\centering
		
		\begin{tabular}{@{ }c@{ }|@{ }c@{ }|@{ }c@{ }|c}
			\hline
			\textbf{Datasets} & \textbf{Attacks} & \textbf{\# Test Pkts (Malicious)} & \textbf{\# Training Pkts} \\ \hline
			\multirow{3}{*}{\begin{tabular}[c]{@{}c@{}}Kitsune\\ Dataset\end{tabular}} & Mirai Botnet & 10,000 (8,079) & \multirow{6}{*}{\begin{tabular}[c]{@{}c@{}}100,000\\(50,000 benign \&\\50,000 malicious)\end{tabular}} \\ 
			& Fuzzing & 20,000 (14,898) &  \\ 
			& SSDP DoS & 10,000 (7,987) &  \\ \cline{1-3}
			\multirow{3}{*}{\begin{tabular}[c]{@{}c@{}}CIC\\ IDS2017\end{tabular}} & Port Scan & 10,000 (2,569) &  \\ 
			& Brute Force & 20,000 (6,136) &  \\ 
			& DDoS & 10,000 (9,966) &  \\  
			\hline
		\end{tabular}
	\end{table}
	
	\begin{table}[]
		\caption{Notations in Metrics.}	
		\label{tab7}
		\footnotesize
		\centering
		\begin{tabular}{@{}c@{}|l}
			\hline
			\textbf{Notation} & \multicolumn{1}{c}{\textbf{Meaning}} \\ \hline
			$\mathrm{Pos}$ & predicted positive number in original malicious traffic  \\ \hline
			\rule{0pt}{9pt}
			$\widehat{\mathrm{Pos}}$ & predicted positive number in mutated malicious traffic \\ \hline
			\rule{0pt}{9pt}
			\begin{tabular}[c]{@{}c@{}}$\widehat{\mathrm{Pos}}_{mal},$\\ $\widehat{\mathrm{Pos}}_{craft}$\end{tabular}
			&
			\begin{tabular}[c]{@{}l@{}}malicious and crafted pkts' number in $\widehat{\mathrm{Pos}}$ \\ ($\widehat{\mathrm{Pos}}=\widehat{\mathrm{Pos}}_{mal}+\widehat{\mathrm{Pos}}_{craft}$)\end{tabular}
			\\ \hline
			$\mathbf{F}_{adver}$& the set of adversarial features \\ \hline
			$\mathbf{F}_{mal}$ & the set of features extracted from original malicious traffic \\ \hline
			\rule{0pt}{9pt}
			$\mathbf{\hat{F}}_{mal}$ & the set of features extracted from mutated malicious traffic \\ \hline
			$\mathbb{E}(\cdot)$& mathematical expectation \\ \hline
		\end{tabular}
	\end{table}
	
	\begin{table*}[!t]
		\caption {Experimental Metrics.}
		\label{tab6}
		\small
		\centering
		\renewcommand\arraystretch{1.1}
		\begin{tabular}{@{}c@{}|@{}c@{}|@{ }p{6cm}@{ }}
			\hline
			\textbf{Metric Name} & \textbf{Formulation} & \multicolumn{1}{@{ }c}{\textbf{Intuitive Description}} \\ \hline
			\textbf{D}etection \textbf{E}vasion \textbf{R}ate (\textbf{DER}) & $1-\frac{\widehat{\mathrm{Pos}}}{\mathrm{Pos}}$ & \tabincell{p{6cm}}{\% undetected mutated traffic (malicious and crafted) to originally detectable traffic.} \\ \cline{1-3} 
			
			original \textbf{M}alicious traffic \textbf{E}vasion \textbf{R}ate (\textbf{MER}) & $1-\frac{\widehat{\mathrm{Pos}}_{mal}}{\mathrm{Pos}}$ & \tabincell{p{6cm}}{\% undetected mutated malicious traffic (exclude crafted) to originally detectable traffic.} \\ \cline{1-3} 
			
			malicious \textbf{P}robability \textbf{D}ecline \textbf{R}ate (\textbf{PDR}) & $1-
			\mathbb{E}_{(\boldsymbol{f},\hat{\boldsymbol{f}}) \in (\mathbf{F}_{mal},\mathbf{\hat{F}}_{mal})}
			\frac{\mathcal{C}(\hat{\boldsymbol{f}})}{\mathcal{C}(\boldsymbol{f})}
			$ & \tabincell{p{6cm}}{To which extent the malicious probability output declines in the targeted ML classifier.} \\ \hline
			
			\textbf{M}alicious features \textbf{M}imicry \textbf{R}ate (\textbf{MMR}) &
			$1-\mathop{\mathbb{E}}_{\substack{(\boldsymbol{f},\hat{\boldsymbol{f}},\boldsymbol{f}_{a}) \in (\mathbf{F}_{mal},\mathbf{\hat{F}}_{mal},\mathbf{F}_{adver})}}
			\frac{\mathcal{L}(\hat{\boldsymbol{f}},\boldsymbol{f}_{a})}
			{\mathcal{L}(\boldsymbol{f},\boldsymbol{f}_{a})}$ 
			& \tabincell{p{6cm}}{To which extent features extracted from mutated traffic are close to the adversarial features.} \\ \hline
		\end{tabular}
	\end{table*}

	\begin{figure*}[]
		\centering
		\includegraphics[width=\linewidth]{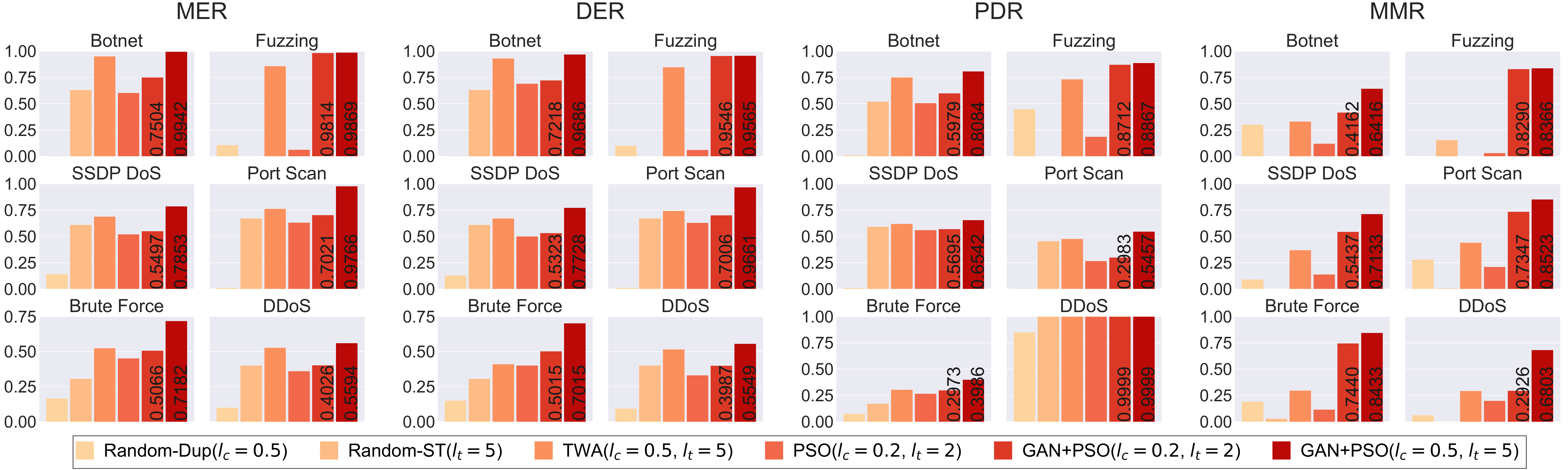}
		\caption{The evasive effectiveness of our attacks compared with baselines (\textbf{\emph{higher is better from the perspective of attacks}}).}
		\label{fig7}
	\end{figure*}

	\noindent \textbf{Metrics}.
	We firstly present four new metrics with formulations and intuitive descriptions listed in Table \ref{tab6}. Notations used in the metrics are listed in Table \ref{tab7} (Higher is better for all metrics).
	According to usage, evaluation metrics used in this work can be divided into three categories:
	\begin{itemize}
		\item \textbf{Evasive effectiveness} (MER, DER, and PDR). 
		\emph{Original Malicious traffic Evasion Rate (MER)} and \emph{Detection Evasion Rate (DER)} respectively reflect how much \emph{original malicious traffic} and \emph{all mutated traffic} (including crafted traffic) become evasive. That is to say, DER additionally considers whether the crafted traffic is classified as malicious, which can reflect whether our attack is stealthy.
		Since MER and DER are highly dependent on the anomaly threshold, we propose a more accurate metric by measuring the decline rate of the malicious probabilities outputted by the targeted classifier, namely \emph{malicious Probability Decline Rate (PDR)}.
		The relationship between evasive effectiveness of attacks and robustness of NIDSs is: \textbf{For the same NIDS, attacks with better evasive effectiveness are more useful for evaluating robustness. For the same attack, NIDSs under better evasive effectiveness are less robust.} 
		
		\item \textbf{Interpretable indicator} (MMR). In order to explain and  understand the reason and principle of evasion attacks on ML-based NIDSs, we propose an interpretable indicator \emph{Malicious features Mimicry Rate (MMR)} which can explicitly show the change of features in the latent space during attacks. Specifically, MMR reflects the degree to which malicious features are close to adversarial features during the mutation.

		\item \textbf{Detection performance}. We additionally use three typical metrics---\emph{Precision}, \emph{Recall}, and \emph{F1-score}---to measure the detection performance of NIDSs. 	
	\end{itemize}

	\subsection{Evasive Effectiveness of Different Attacks}
	\label{sec7.2}
	In this section, we compare the evasive effectiveness of our PGA attacks with three baselines by evading \texttt{Kitsune} under different traffic sets.
	We also evaluate the effectiveness of adversarial features in our attacks by comparing our PSO-based algorithm with ({\small\textsf{GAN+PSO}}) and without ({\small\textsf{PSO}}) adversarial features. {\small\textsf{PSO}} only uses benign features (instead of adversarial features) when evaluating fitness (on line 6 of Alg. \ref{alg1}).
	We also compare the impact of overhead budget in our attacks using a lower ($l_{c}=0.2,l_{t}=2$) and a higher budget ($l_{c}=0.5,l_{t}=5$). Baseline attacks are all with the higher overhead budget. The results are shown in Fig. \ref{fig7}.
	
	\noindent \textbf{Evasive effectiveness comparison}. As evident in the results of MER/PDR, our attack {\small\textsf{GAN+PSO}} performs very well relative to random mutations at the same budget ($l_{c}=0.5,l_{t}=5$). The effectiveness of random mutations is extremely unstable; each mutation only works under specific traffic sets.
	Moreover, our attack outperforms the state-of-the-art white-box attack ({\small\textsf{TWA}}), which also demonstrates the superiority of our attack method since the performance of white-box attacks is generally supposed to be better. 
	As for DER, results show that the drop from MER to DER is <3\% in most cases, which shows that the crafted traffic in our evasion attack is stealthy and unobservable even with $l_{c}=0.5$ (more crafted packets).
	
	\noindent \textbf{Impact of adversarial features}. We observe that using adversarial features indeed increases the evasive effectiveness (by 10-20\% usually). Especially in {\small\emph{\textsf{Fuzzing}}}, {\small\textsf{GAN+PSO}} increases MER/DER by more than 90\% compared with {\small\textsf{PSO}}.
	
	\noindent \textbf{Impact of overhead budget}. It is easy to understand that a higher overhead budget (namely, looser limitation) performs better results. Specifically, 
	{\small\textsf{GAN+PSO}} with larger $l_{c}$ and $l_{t}$ have 20-30\% higher MER/DER in most cases.
	
	\begin{table*}[!t]  
		\caption{The evasive effectiveness on NIDSs with other feature extractors and ML classifiers.}
		\label{tab5}
		
		\footnotesize
		\renewcommand\arraystretch{0.9} 
		\setlength\tabcolsep{0.8ex}
		
		\begin{minipage}[]{0.09\textwidth}
			\centering
			\begin{threeparttable}
				
				\vspace{3.55mm}
				\begin{tabular}{c|} 
					\toprule  
					\multirow{2}*{\begin{tabular}[c]{@{ }c@{ }} Feature \vspace{1ex}\\ Extractor\end{tabular}} \\
					\vspace{1.4ex}

					\\
					\midrule
					
					\multirow{6}*{\textbf{AfterImage}}
					\\
					\\
					\\
					\\
					\\
					\\
					
					\cmidrule(lr){1-1}
					
					\multirow{6}*{\begin{tabular}[c]{@{ }c@{ }}\textbf{CIC} \\ \textbf{FlowMeter}\end{tabular}}
					\\
					\\
					\\
					\\
					\\
					\\
					\bottomrule  
				\end{tabular}  
			\end{threeparttable}
		\end{minipage}
		\hspace{1.5ex}
		\begin{minipage}[]{0.43\textwidth}
			\renewcommand\arraystretch{0.9} 
			\begin{threeparttable}
				\centerline {(a) \emph{\textsf{Botnet}}}
				\vspace{1mm}
				\begin{tabular}{@{ }c@{ }ccccccc} 
					\toprule  
					\multirow{2}*{\begin{tabular}[c]{@{ }c@{ }} ML \vspace{1ex} \\ Classifier\end{tabular}}&
					\multicolumn{3}{c}{Detection}&\multicolumn{4}{c}{ Evasive (\textbf{MER})---\textbf{\emph{higher is better}}} \\
					\cmidrule(lr){2-4} \cmidrule(lr){5-8} 
					&{P}&{R}&{F1}
					&{\textsf{R-Dup}}&{\textsf{R-ST}}&{\textsf{TWA}}&{\textbf{Ours}}\\
					\midrule
					
					KitNET&0.98&0.92&0.95&	0.20\%&63.28\%&94.98\%&\textbf{99.42\%} \\
					LR&0.96&0.90&0.93&		0.67\%&14.96\%&50.17\%&\textbf{54.74\%} \\
					DT&0.79&0.90&0.84&		0.61\%&14.36\%&49.13\%&\textbf{60.36\%} \\
					SVM&0.99&0.90&0.94&		0.82\%&9.10\%&32.59\%&\textbf{40.31\%} \\
					MLP&0.96&0.97&0.97&		0.87\%&4.72\%&10.59\%&\textbf{45.15\%} \\
					IF&0.95&0.93&0.94&		0.76\%&0.16\%&0.52\%&\textbf{33.63\%} \\
					
					\cmidrule(lr){1-8}
					
					KitNET&0.87&0.98&0.92&	0.00\%&9.69\%&29.31\%&\textbf{38.89\%} \\
					LR&0.79&0.97&0.87&		2.48\%&1.87\%&20.37\%&\textbf{40.74\%} \\
					DT&0.76&0.91&0.83&		0.64\%&3.70\%&17.90\%&\textbf{30.76\%} \\
					SVM&0.78&0.98&0.87&		0.00\%&22.19\%&41.35\%&\textbf{84.62\%} \\
					MLP&0.90&0.88&0.89&		0.00\%&0.00\%&9.10\%&\textbf{38.80\%} \\
					IF&0.97&0.89&0.93&		2.46\%&0.00\%&0.00\%&\textbf{37.31\%} \\
					\bottomrule  
				\end{tabular}  
			\end{threeparttable}
		\end{minipage}
		\hspace{0.5ex}
		\begin{minipage}[]{0.43\textwidth}
			\renewcommand\arraystretch{0.9} 
			\begin{threeparttable}
				\centerline {(b) \emph{\textsf{DDoS}}}
				\vspace{1mm}
				\begin{tabular}{@{ }c@{ }ccccccc} 
					\toprule  
					\multirow{2}*{\begin{tabular}[c]{@{ }c@{ }} ML \vspace{1ex} \\ Classifier\end{tabular}}&
					\multicolumn{3}{c}{Detection}&\multicolumn{4}{c}{ Evasive (\textbf{MER})---\textbf{\emph{higher is better}}} \\
					\cmidrule(lr){2-4} \cmidrule(lr){5-8} 
					&{P}&{R}&{F1}
					&{\textsf{R-Dup}}&{\textsf{R-ST}}&{\textsf{TWA}}&{\textbf{Ours}}\\
					\midrule
					
					KitNET&0.94&0.97&0.95&	10.41\%&13.74\%&52.75\%&\textbf{55.94\%}\\
					LR&0.96&0.91&0.93&		27.09\%&16.49\%&64.90\%&\textbf{70.59\%}\\
					DT&0.76&0.91&0.83&		27.43\%&17.43\%&64.27\%&\textbf{69.86\%}\\
					SVM&0.99&0.90&0.94&		29.96\%&18.20\%&35.19\%&\textbf{79.55\%}\\
					MLP&0.98&0.91&0.94&		25.04\%&9.58\%&43.47\%&\textbf{50.63\%}\\
					IF&0.85&0.89&0.87&		0.00\%&12.99\%&0.0\%&\textbf{17.71\%}\\
					
					\cmidrule(lr){1-8}
					
					KitNET&0.93&0.90&0.92&	0.00\%&8.77\%&28.59\%&\textbf{32.04\%}\\
					LR&0.70&0.73&0.71&		1.25\%&0.00\%&14.80\%&\textbf{36.82\%}\\
					DT&0.67&0.73&0.70&		0.00\%&3.01\%&16.90\%&\textbf{35.56\%}\\
					SVM&0.75&0.74&0.74&		4.13\%&2.04\%&35.75\%&\textbf{40.92\%}\\
					MLP&0.72&0.71&0.72&		5.58\%&15.45\%&42.89\%&\textbf{50.35\%}\\
					IF&1.00&0.89&0.94&		0.00\%&0.00\%&13.44\%&\textbf{25.98\%}\\
					\bottomrule  
				\end{tabular}  
			\end{threeparttable}
		\end{minipage}
	\end{table*}

	\begin{table*}[]  
		\caption{The evasive effectiveness of our attacks with limited knowledge of features (PBAs) compared with PGA.}
		\label{tab10}
		
		\footnotesize
		
		\renewcommand\arraystretch{0.925} 
		\setlength\tabcolsep{1.2ex}
		\centering
		\begin{threeparttable}    
			\begin{tabular}{ccccccc} 
				\toprule  
				\multirow{2}*{\begin{tabular}[c]{@{ }c@{ }} Attacks \vspace{1ex} \\ (Knowledge on features)\end{tabular}}&  
				\multicolumn{6}{c}{The Evasive Effectiveness (\textbf{MER} / \textbf{PDR}) on Different Traffic Sets---\textbf{\emph{higher is better}}} \\ 
				\cmidrule(lr){2-7}
				&\emph{\textsf{Botnet}}&\emph{\textsf{Fuzzing}}&\emph{\textsf{SSDP}}&\emph{\textsf{Port Scan}}&\emph{\textsf{Brute Force}}&\emph{\textsf{DDoS}}\\   
				\midrule  
				PBA(0\hspace{1ex}\%) 	
				&83.46\% / 68.47\%	&82.68\% / 70.04\%
				&53.19\% / 57.31\% &35.56\% / 26.32\%
				&49.50\% / 28.07\% &33.06\% / 99.99\%  \\
				PBA(50\%) 	&98.77\% / 76.16\%	&98.64\% / 82.38\%
				&68.79\% / 62.16\%	&72.72\% / 49.78\%
				&52.43\% / 30.39\%	&41.22\% / 99.99\%  \\
				PBA(75\%) 	&99.28\% / 77.87\%	&98.26\% / 81.89\%
				&82.62\% / 67.06\%	&76.82\% / 52.38\%
				&60.12\% / 33.48\%	&50.45\% / 99.99\% \\
				\cdashline{1-7}[3pt/2pt]
				\rule{0pt}{7pt}PGA(100\%) 
				&99.42\% / 80.84\%	&98.69\% / 88.67\%
				&78.53\% / 65.42\%	&97.66\% / 54.57\%
				&71.81\% / 39.86\%	&55.94\% / 99.99\%  \\
				\bottomrule  
			\end{tabular}  
		\end{threeparttable}
	\end{table*}

	\noindent \textbf{Performance of different traffic sets}. As shown in the results, our attack achieves >97\% MER/DER on half of the traffic sets, as well as >70\% MER/DER on five of six  traffic sets. As for reasons of the relatively poor MER/DER in {\small\emph{\textsf{DDoS}}}, this is because malicious features are originally farther from the benign space and beyond the attacker's ability/budget (See Scenario 3 in Figure \hyperref[fig2b]{2b}, if a malicious feature is beyond the attacker's overhead budget/ability, it cannot be transformed into a benign one by any means). 
	In fact, we find the anomaly score (i.e, RMSE in \texttt{Kitsune}) of original features in {\small\emph{\textsf{DDoS}}} is many orders of magnitude larger than other scenarios. This is exactly why its PDR is higher than others (over 99.99\%) but MER/DER is lower.
	This finding also shows that it is necessary to consider attacker's ability/budget on mutating traffic as well as the original intensity of anomaly instead of purely comparing the evasion rate. 
	
	\subsection{Robustness of Different Classifiers and Features}
	\label{sec7.3}
	We conduct evasion attacks (our PGA and baselines with the higher budget) on different NIDSs described in Section \ref{sec7.1} under {\small\emph{\textsf{Botnet}}} and {\small\emph{\textsf{DDoS}}} traffic (The other four traffic sets are not shown due to the reason of space, and using the current two sets is enough to draw the same conclusions). Since DER has been found to be very similar to MER, we use MER to measure the evasive performance, which is also the most concerning indicator for attackers. PDR is not used due to the inconsistency among ML models. Table \ref{tab5} lists the results.

	\noindent \textbf{Evasive effectiveness comparison}. Compared with baseline attacks (each row in the table), our attack has broader generality for evading various ML classifiers using different kinds of features. Specifically, {\small\textsf{Random-RT}} always performs poorly while {\small\textsf{Random-Dup}} only has evasive effectiveness for a few cases. Once again, our attack outperforms {\small\textsf{TWA}} in all cases, especially for the Isolation Forest model. We attribute the generality to the feature-level mimicking in our model-agnostic attack.
	Meanwhile, it can be observed that the evasive performance of different classifiers is diverse significantly (each column of MER in the table). 
	We think this is sound since the robustness of various models is different (just as their detection performance is also different). 
	
	\noindent \textbf{Robustness of different feature sets}. NIDSs with flow-based features are slightly more robust against our attack as well as other attacks than packet-based ones. 
	Note that, our mutation operators are packet-based (but not flow-based) due to the generic black/gray-box assumption. Per-packet mutation can attack flow-based NIDSs since flows consist of packets.
	
	\noindent \textbf{Robustness of different classifiers}. Based on the results in {\small\emph{\textsf{Botnet}}}, we find that traditional ML methods are more robust than deep neural networks. Specifically, KitNET (in \texttt{Kitsune}) has the (almost) best detection performance but also suffers the highest evasion rate. In addition to the inherent vulnerability of neural networks\cite{szegedy_intriguing_2014}, another probable reason is that KitNET clusters the features into groups, which gives attackers a better chance to influence more feature groups. Through experiments, we find the top 10\% dimensions in original features exploited by our attack can eventually cover more than 50\% of the features groups. 
	The robustness of the different methods in {\small\emph{\textsf{DDoS}}} is poor, while only IF still maintains good robustness.
	From the results of all cases, IF is relatively more robust, while the robustness of other ML models is unstable.

	\subsection{Attacks with Limited Knowledge of Features}
	\label{sec7.4}
	So far, we have evaluated the effectiveness of our attack under the PGA assumption. We now extend our attack with limited knowledge of features used in targeted NIDSs (i.e., PBA).
	Specifically, we evaluate three types of attackers, who know the 
	75\%, 50\%, and 0\% features that are accurately used by NIDSs, respectively. Recall Section \ref{sec3.1}, the only difference between PBA and PGA is the surrogate feature extractor used by the attacker. For PBA, besides limited known features, the attacker also extracts all commonly used features in Appendix \ref{app1}. Especially for PBA(0\%), the attacker without any knowledge about the target system can only simulate the extractor by using other features. We use \texttt{Kitsune} as the targeted NIDS and evaluate the MER and PDR of three PBAs and the PGA in all traffic sets. The results are in Table \ref{tab10}.
	
	As mentioned before, we think that PDR can better reflect the evasive effectiveness compared with MER. As shown in the results, PBA(50\%) and PBA(75\%) perform high PDR that is similar to PGA. Even for attackers without any knowledge, PBA(0\%) still has a strong evasive ability (Compared with PGA, the drop of PDR is within 20\%). The key insight is that even if we cannot accurately know the features used by NIDSs, the mutation method computed by our attack through simulated features is also effective on real features. This finding seems to be very frustrating and frightening for such ML-based NIDSs, meaning that a weak attacker can easily make a considerable portion of malicious traffic become evasive.
	
	\begin{figure}[]
		\centering
		\includegraphics[width= 0.99\linewidth]{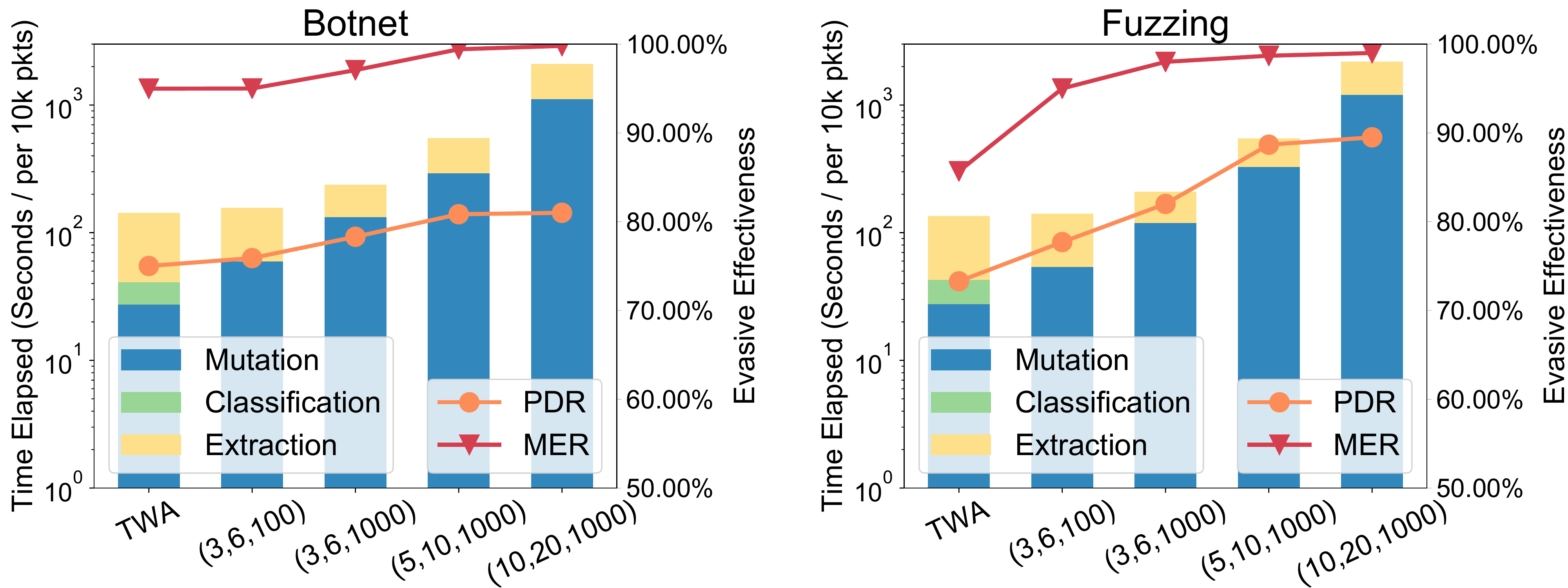}
		\caption{Execution Cost and Impact of Parameters}
		\label{fig11}
	\end{figure}

	\subsection {Execution Cost and Impact of Parameters}
	\label{sec7.5}
	It is necessary to measure the execution cost for attacks, especially for attackers with limited computing resources. Here, we use TWA as a comparison algorithm to represent the \emph{lower bound} of execution time. This is because TWA requires the output value of the classifier, so the quality of candidate solutions (i.e., traffic mutants) can be quickly measured, but we do not have this knowledge in our attack. We also compare the impact of three key parameters on the execution time and evasive performance. We denote our attacks using  different parameters with {\small\textsf{($N_{iter}$,$N_{swarm}$,$N_{adver}$)}}: $N_{iter}$ and $N_{swarm}$ denotes the number of iterations and particles in PSO, and $N_{adver}$ denotes the number of adversarial features. The results are shown in Fig. \ref{fig11}, where we use \texttt{Kitsune} under two traffic sets since other sets have similar results.
	
	It is shown that our attack with {\small\textsf{(3,6,100)}} can approximate the execution time of TWA while performing better evasive performance than TWA. For other parameters, our attacks are acceptable in execution time. Larger parameters have better evasive performance but will consume more time. To balance this trade-off, we think {\small\textsf{(5,10,1000)}} is the best combination of parameters, which is also chosen for other parts of experiments.

	\subsection{Verification of Malicious Functionality}
	\label{sec7.6}
	To be rigorous, although we guarantee the mutation operators in Section \ref{sec5.2} will not compromise the malicious functionality of original traffic, we still verify the malicious functionality of the mutated traffic in all six traffic sets.
	
	To measure the malicious functionality, we use three types of indicators: \emph{attack effect}, \emph{malicious behavior} and \emph{attack efficiency}, and compare them in original and mutated traffic. Take {\small\emph{\textsf{Botnet}}} as an example to illustrate the three indicators: In our selected traffic, an attacker used a malware called Mirai to scan IoT devices in the LAN and successfully scanned 8 open devices. In this scenario, \emph{attack effect} is the final result of the attack, which is that 8 devices were successfully scanned. 
	\emph{Malicious behavior} contains all offensive behaviors regardless of whether they eventually affect, which is the number of scans. \emph{Attack efficiency} is related to the elapsed time of the attack. Obviously, \emph{attack effect} has a greater impact on functionality than \emph{malicious behavior}, and the change rate of \emph{attack efficiency} must be within the attacker's overhead budget $l_{t}$.

	We use VMs and Dockers to simulate the experimental testbed for each traffic set by referring to their papers \cite{mirsky_kitsune:_2018,sharafaldin_toward_2018}.
	Then we use Tcpreplay and Tcplivereplay to replay original and mutated attack traffic in the testbed and observe the three indicators. Since different attacks have diverse functionalities, the specific meanings of three indicators are case-by-case, making the validation experiment straightforward but tedious, so we put the details in Appendix \ref{app22} and leave the result here: \emph{Mutated traffic generated through our evasion attack can preserve the malicious functionality}. Specifically, \emph{attack effect} keeps unchanged in all cases and \emph{malicious behavior} is reduced only in DoS/DDoS attacks (attack bandwidth is decreased due to increased time interval, but this reduction does not exceed the attacker's budget ($l_{t}$). As for \emph{attack efficiency}, although our method may slow down some kinds of attack, the change rate of elapsed time is always $< l_{t}$.
	
	\begin{figure}[]
		\centering
		\includegraphics[width= 0.965\linewidth]{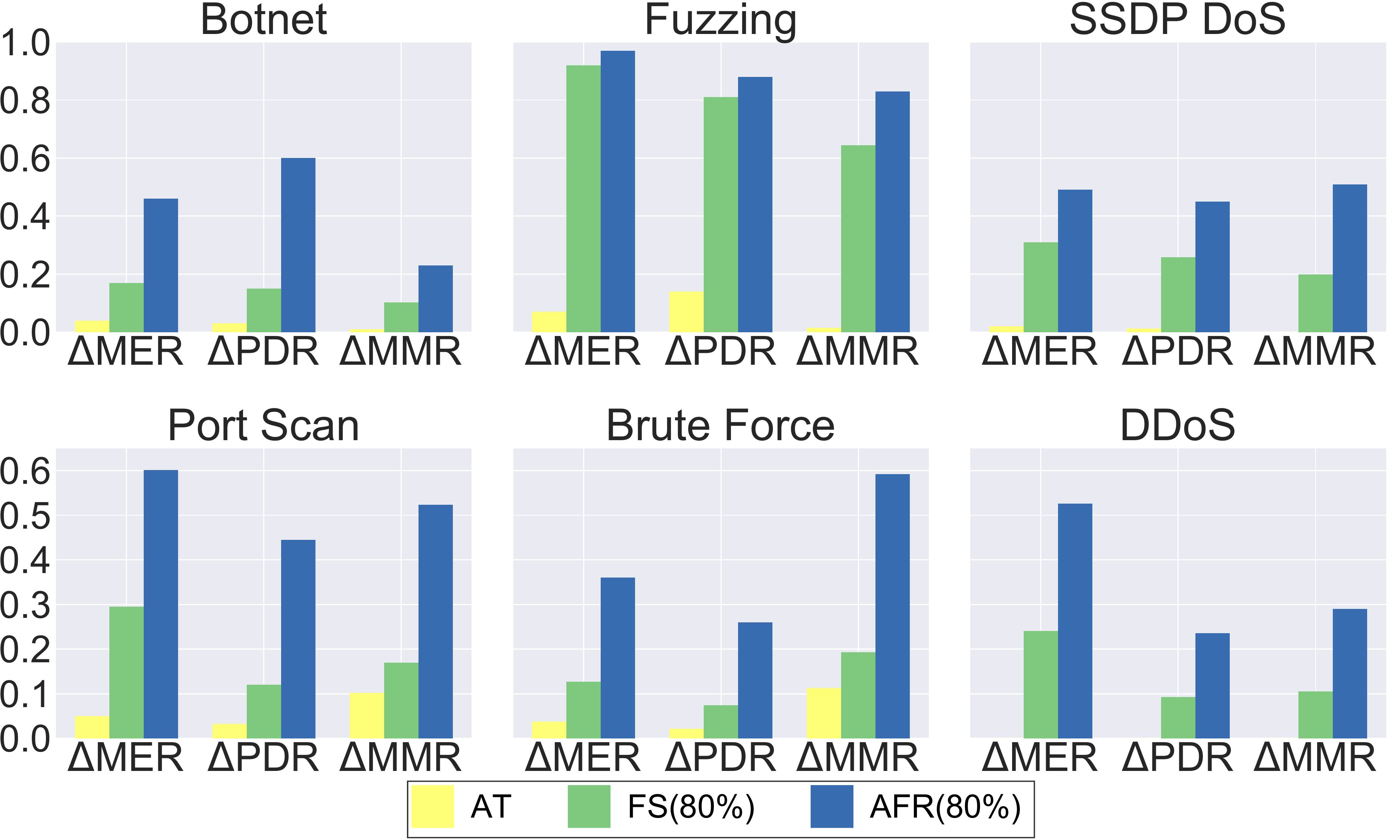}
		\caption{The defense performance (\textbf{\emph{higher is better}}).}
		\label{fig8}
	\end{figure}
	
	\subsection{Performance of Defense Schemes}
	\label{sec7.7}
	In this section, we evaluate three defenses mentioned in Section \ref{sec6}.
	For adversarial training ({\small\textsf{AT}}), we retrain the ML classifiers with 80\% relabeled adversarial features and use the remaining 20\% for testing. For feature selection ({\small\textsf{FS}}), we use embedded Lasso regression model to retain 80\% dimensions. As for our adversarial feature reduction ({\small\textsf{AFR}}), we also retain 80\% feature dimensions. We use \texttt{Kitsune} in all traffic sets, and use the decline of metrics ($\Delta$MER/PDR/MMR, also can be viewed as the improvement of robustness) to evaluate the defense performance. The results are shown in Fig. \ref{fig8}.
	
	Compared with {\small\textsf{AT}} and {\small\textsf{FS}}, our {\small\textsf{AFR}} is very effective for improving the robustness of NIDSs by reducing MER/PDR/MMR.
	We observe that {\small\textsf{AT}} has a very limited and unstable defensive effect against the proposed attack. This is because it can limit the generation of adversarial features, but cannot prevent vulnerable feature dimensions from being exploited during traffic mutation. {\small\textsf{FS}} can perform better defense effectiveness in some cases. This shows that using fewer feature dimensions can increase the difficulty for attackers to transform the entire malicious features to some extent.
	We also measure the change of F1-score to evaluate whether the defense methods compromise the original detection performances. 
	For reasons of spaces, they are not depicted since the change is quite small (within $\pm$5\%).

	\section{Discussions}
	\label{sec8}
	We discuss limitations, considerations and potential improvements of our attacks as follows.
	
	\noindent \textbf{Limitations.} As mentioned before, our method is designed for evading NIDSs without payload inspection, so it is invalid for systems additionally using payload-based detection. However, this problem can be easily solved by combining the \emph{polymorphic blending attack} \cite{fogla_polymorphic_2006} with ours. And this can be easily implemented: leveraging polymorphic blending attack to encrypt the payload of original malicious traffic and using our method to inject crafted packets. Another limitation is that our attack is offline at present, but this can be solved by replaying mutated traffic since we have proofed that replayed traffic can conduct the same malicious intent as the original attack.  
	
	\noindent \textbf{Why black-box attacks.} Compared with white-box attacks, black-box attacks are with higher feasibility since they need fewer prerequisites/requirements, especially in practical  settings\cite{papernot2017practical}. Note that, our scope in this work is from the perspective of attackers to propose more practical adversarial attacks and then propose the corresponding defense. Future work can focus on designing theoretically secure systems from the perspective of defense (according to Kerckhoff's principle \cite{kerckhoffs1883cryptographic}).
	
	\noindent \textbf{Background traffic.} In the proposed attack, we inject some crafted traffic which can be aggregated with original packets in order to impact features. However, some unpredictable background traffic (i.e., some traffic that is not controlled by the attacker but can also reach the victim or NIDS) may disrupt our mutated traffic on some features. Nonetheless, we find that only features aggregated by destination information (e.g., dstIP) are affected. Thus, the impact of background traffic is extremely limited (e.g., \texttt{Kitsune} has no features extracted only by destination). 
	
	\noindent \textbf{Improving the attack effect.} In this paper, we pay more attention to explore a more practical attack rather than try our best to improve the evasion rate. 
	For one thing, we only use the default settings in the implementation of PSO and GAN in this study. For example, we simply use the parameters of PSO algorithm recommended in \cite{eberhart_comparing_2000}.
	For another, we use Euclidean distance to measure the similarity of features in this work.
	Future work should focus on whether other distance function or careful parameter tuning can perform better results.

	\section{Conclusion}
	\label{sec10}
	
	This paper describes the first step toward developing a systematic study on practical traffic-space evasion attacks for adversarial robustness evaluation on ML-based NIDSs.
	Experimental results show our attack is effective (>97\% evasion rate in half cases) and the proposed defense method can effectively mitigate such attacks.
	Our attack outperforms prior works while using affordable execution cost, and is effective even without any knowledge of the targeted systems.
	We extensively measure the robustness of various ML-based NIDSs and provide important findings.
	Our finding demonstrates that the paradigm of feature engineering should be shifted; we deem the detection performance together with anti-evasion robustness both need to be taken into consideration while designing feature sets.
	We firmly believe that our work provides important insights for improving the robustness of ML-based NIDSs and inspires more attention to the robust feature engineering in all ML-based systems.

	
	\label{bibref}
	\bibliographystyle{ieeetr}
	\bibliography{ref}

\begin{thebibliography}{10}

\bibitem{garcia2009anomaly}
P.~Garcia-Teodoro, J.~Diaz-Verdejo, G.~Maci{\'a}-Fern{\'a}ndez, and
  E.~V{\'a}zquez, ``Anomaly-based network intrusion detection: Techniques,
  systems and challenges,'' {\em computers \& security}, vol.~28, no.~1-2,
  pp.~18--28, 2009.

\bibitem{hodo_shallow_2017}
E.~Hodo, X.~Bellekens, A.~Hamilton, C.~Tachtatzis, and R.~Atkinson, ``Shallow
  and deep networks intrusion detection system: {A} taxonomy and survey,'' {\em
  arXiv preprint arXiv:1701.02145}, 2017.

\bibitem{gulghane2019survey}
S.~Gulghane, V.~Shingate, S.~Bondgulwar, G.~Awari, and P.~Sagar, ``A survey on
  intrusion detection system using machine learning algorithms,'' in {\em
  International Conference on Innovative Data Communication Technologies and
  Application}, pp.~670--675, Springer, 2019.

\bibitem{mirsky_kitsune:_2018}
Y.~Mirsky, T.~Doitshman, Y.~Elovici, and A.~Shabtai, ``Kitsune: an ensemble of
  autoencoders for online network intrusion detection,'' {\em Network and
  Distributed System Security Symposium (NDSS)}, 2018.

\bibitem{zhong2020helad}
Y.~Zhong, W.~Chen, Z.~Wang, Y.~Chen, K.~Wang, Y.~Li, X.~Yin, X.~Shi, J.~Yang,
  and K.~Li, ``Helad: A novel network anomaly detection model based on
  heterogeneous ensemble learning,'' {\em Computer Networks}, vol.~169,
  p.~107049, 2020.

\bibitem{doriguzzi2020lucid}
R.~Doriguzzi-Corin, S.~Millar, S.~Scott-Hayward, J.~Martinez-del Rincon, and
  D.~Siracusa, ``Lucid: A practical, lightweight deep learning solution for
  ddos attack detection,'' {\em IEEE Transactions on Network and Service
  Management}, vol.~17, no.~2, pp.~876--889, 2020.

\bibitem{xu2020method}
C.~Xu, J.~Shen, and X.~Du, ``A method of few-shot network intrusion detection
  based on meta-learning framework,'' {\em IEEE Transactions on Information
  Forensics and Security}, vol.~15, pp.~3540--3552, 2020.

\bibitem{ferrag2020deep}
M.~A. Ferrag, L.~Maglaras, S.~Moschoyiannis, and H.~Janicke, ``Deep learning
  for cyber security intrusion detection: Approaches, datasets, and comparative
  study,'' {\em Journal of Information Security and Applications}, vol.~50,
  p.~102419, 2020.

\bibitem{barreno_security_2010}
M.~Barreno, B.~Nelson, A.~D. Joseph, and J.~D. Tygar, ``The security of machine
  learning,'' {\em Machine Learning}, vol.~81, pp.~121--148, 2010.

\bibitem{tygar_adversarial_2011}
J.~D. Tygar, ``Adversarial {Machine} {Learning},'' {\em IEEE Internet
  Computing}, vol.~15, pp.~4--6, 2011.

\bibitem{szegedy_intriguing_2014}
C.~Szegedy, W.~Zaremba, I.~Sutskever, J.~Bruna, D.~Erhan, I.~J. Goodfellow, and
  R.~Fergus, ``Intriguing properties of neural networks,'' {\em International
  Conference on Learning Representations (ICLR)}, 2014.

\bibitem{carlini_towards_2017}
N.~Carlini and D.~A. Wagner, ``Towards {Evaluating} the {Robustness} of
  {Neural} {Networks},'' {\em IEEE Symposium on Security and Privacy (S\&P)},
  pp.~39--57, 2017.

\bibitem{goodfellow_explaining_2015}
I.~J. Goodfellow, J.~Shlens, and C.~Szegedy, ``Explaining and {Harnessing}
  {Adversarial} {Examples},'' in {\em International {Conference} on {Learning}
  {Representations} ({ICLR})}, 2015.

\bibitem{li_stealthy_2019}
S.~Li, A.~Neupane, S.~Paul, C.~Song, S.~V. Krishnamurthy, A.~K.~R. Chowdhury,
  and A.~Swami, ``Stealthy {Adversarial} {Perturbations} {Against}
  {Real}-{Time} {Video} {Classification} {Systems},'' in {\em Network and
  {Distributed} {System} {Security} {Symposium} ({NDSS})}, 2019.

\bibitem{li_textbugger:_2019}
J.~Li, S.~Ji, T.~Du, B.~Li, and T.~Wang, ``{TextBugger}: {Generating}
  {Adversarial} {Text} {Against} {Real}-world {Applications},'' in {\em Network
  and {Distributed} {System} {Security} {Symposium} ({NDSS})}, 2019.

\bibitem{laskov_practical_2014}
P.~Laskov and {others}, ``Practical evasion of a learning-based classifier: {A}
  case study,'' in {\em {IEEE} {Symposium} on {Security} and {Privacy}
  ({S}\&{P})}, pp.~197--211, IEEE, 2014.

\bibitem{xu_automatically_2016}
W.~Xu, Y.~Qi, and D.~Evans, ``Automatically evading classifiers,'' in {\em
  Network and {Distributed} {System} {Security} {Symposium} ({NDSS})}, 2016.

\bibitem{hu_generating_2017}
W.~Hu and Y.~Tan, ``Generating adversarial malware examples for black-box
  attacks based on {GAN},'' {\em arXiv preprint arXiv:1702.05983}, 2017.

\bibitem{chen2020training}
Y.~Chen, S.~Wang, D.~She, and S.~Jana, ``On training robust $\{$PDF$\}$ malware
  classifiers,'' in {\em 29th $\{$USENIX$\}$ Security Symposium ($\{$USENIX$\}$
  Security 20)}, pp.~2343--2360, 2020.

\bibitem{ptacek_insertion_1998}
T.~H. Ptacek and T.~N. Newsham, ``Insertion, {Evasion}, and {Denial} of
  {Service}: {Eluding} {Network} {Intrusion} {Detection},'' tech. rep., SECURE
  NETWORKS INC CALGARY ALBERTA, Jan. 1998.

\bibitem{handley_network_2001}
M.~Handley, V.~Paxson, and C.~Kreibich, ``Network {Intrusion} {Detection}:
  {Evasion}, {Traffic} {Normalization}, and {End}-to-{End} {Protocol}
  {Semantics},'' in {\em {USENIX} {Security} {Symposium}}, 2001.

\bibitem{cheng_evasion_2012}
T.-H. Cheng, Y.-D. Lin, Y.-C. Lai, and P.-C. Lin, ``Evasion {Techniques}:
  {Sneaking} through {Your} {Intrusion} {Detection}/{Prevention} {Systems},''
  {\em IEEE Communications Surveys \& Tutorials}, vol.~14, pp.~1011--1020,
  2012.

\bibitem{stinson_towards_2008}
E.~Stinson and J.~C. Mitchell, ``Towards {Systematic} {Evaluation} of the
  {Evadability} of {Bot}/{Botnet} {Detection} {Methods}.,'' {\em WOOT}, vol.~8,
  2008.

\bibitem{wright2009traffic}
C.~V. Wright, S.~E. Coull, and F.~Monrose, ``Traffic morphing: An efficient
  defense against statistical traffic analysis.,'' in {\em NDSS}, vol.~9,
  Citeseer, 2009.

\bibitem{homoliak_improving_2018}
I.~Homoliak, M.~Teknos, M.~Ochoa, D.~Breitenbacher, S.~Hosseini, and
  P.~Hanacek, ``Improving {Network} {Intrusion} {Detection} {Classifiers} by
  {Non}-payload-{Based} {Exploit}-{Independent} {Obfuscations}: {An}
  {Adversarial} {Approach},'' {\em arXiv preprint arXiv:1805.02684}, 2018.

\bibitem{hashemi2019towards}
M.~J. Hashemi, G.~Cusack, and E.~Keller, ``Towards evaluation of nidss in
  adversarial setting,'' in {\em Proceedings of the 3rd ACM CoNEXT Workshop on
  Big DAta, Machine Learning and Artificial Intelligence for Data Communication
  Networks}, pp.~14--21, 2019.

\bibitem{liu2016delving}
Y.~Liu, X.~Chen, C.~Liu, and D.~Song, ``Delving into transferable adversarial
  examples and black-box attacks,'' {\em arXiv preprint arXiv:1611.02770},
  2016.

\bibitem{corona_adversarial_2013}
I.~Corona, G.~Giacinto, and F.~Roli, ``Adversarial attacks against intrusion
  detection systems: {Taxonomy}, solutions and open issues,'' {\em Information
  Sciences}, vol.~239, pp.~201--225, Aug. 2013.

\bibitem{chaboya_network_2006}
D.~J. Chaboya, R.~A. Raines, R.~O. Baldwin, and B.~E. Mullins, ``Network
  intrusion detection: {Automated} and manual methods prone to attack and
  evasion,'' {\em IEEE Symposium on Security and Privacy (S\&P)}, 2006.

\bibitem{vigna_testing_2004}
G.~Vigna, W.~Robertson, and D.~Balzarotti, ``Testing network-based intrusion
  detection signatures using mutant exploits,'' in {\em {ACM} {Conference} on
  {Computer} and {Communications} {Security} ({CCS})}, 2004.

\bibitem{mutz_experience_2003}
D.~Mutz, G.~Vigna, and R.~Kemmerer, ``An experience developing an {IDS}
  stimulator for the black-box testing of network intrusion detection
  systems,'' in {\em 19th {Annual} {Computer} {Security} {Applications}
  {Conference}, 2003. {Proceedings}.}, pp.~374--383, IEEE, 2003.

\bibitem{tan_undermining_2002}
K.~M. Tan, K.~S. Killourhy, and R.~A. Maxion, ``Undermining an anomaly-based
  intrusion detection system using common exploits,'' in {\em International
  {Symposium} on {Recent} {Advances} in {Intrusion} {Detection} ({RAID})},
  pp.~54--73, Springer, 2002.

\bibitem{kayacik_mimicry_2008}
H.~G. Kayacik and A.~N. Zincir-Heywood, ``Mimicry attacks demystified: {What}
  can attackers do to evade detection?,'' in {\em 2008 {Sixth} {Annual}
  {Conference} on {Privacy}, {Security} and {Trust}}, pp.~213--223, IEEE, 2008.

\bibitem{kayacik_generating_2009}
H.~G. Kayacik, A.~N. Zincir-Heywood, M.~I. Heywood, and S.~Burschka,
  ``Generating mimicry attacks using genetic programming: a benchmarking
  study,'' in {\em 2009 {IEEE} {Symposium} on {Computational} {Intelligence} in
  {Cyber} {Security}}, pp.~136--143, IEEE, 2009.

\bibitem{fogla_polymorphic_2006}
P.~Fogla, M.~I. Sharif, R.~Perdisci, O.~M. Kolesnikov, and W.~Lee,
  ``Polymorphic {Blending} {Attacks},'' in {\em {USENIX} {Security}
  {Symposium}}, 2006.

\bibitem{fogla_evading_2006}
P.~Fogla and W.~Lee, ``Evading network anomaly detection systems: formal
  reasoning and practical techniques,'' in {\em {ACM} {Conference} on
  {Computer} and {Communications} {Security} ({CCS})}, 2006.

\bibitem{moosavi-dezfooli_deepfool:_2016}
S.-M. Moosavi-Dezfooli, A.~Fawzi, and P.~Frossard, ``{DeepFool}: {A} {Simple}
  and {Accurate} {Method} to {Fool} {Deep} {Neural} {Networks},'' {\em 2016
  IEEE Conference on Computer Vision and Pattern Recognition (CVPR)},
  pp.~2574--2582, 2016.

\bibitem{yuan_adversarial_2019}
X.~Yuan, P.~He, Q.~Zhu, and X.~Li, ``Adversarial examples: {Attacks} and
  defenses for deep learning,'' {\em IEEE transactions on neural networks and
  learning systems}, 2019.

\bibitem{chen2019real}
G.~Chen, S.~Chen, L.~Fan, X.~Du, Z.~Zhao, F.~Song, and Y.~Liu, ``Who is real
  bob? adversarial attacks on speaker recognition systems,'' {\em arXiv
  preprint arXiv:1911.01840}, 2019.

\bibitem{zang2020word}
Y.~Zang, F.~Qi, C.~Yang, Z.~Liu, M.~Zhang, Q.~Liu, and M.~Sun, ``Word-level
  textual adversarial attacking as combinatorial optimization,'' in {\em
  Proceedings of the 58th Annual Meeting of the Association for Computational
  Linguistics}, pp.~6066--6080, 2020.

\bibitem{wang2020adversarial}
Y.~Wang, Y.-a. Tan, W.~Zhang, Y.~Zhao, and X.~Kuang, ``An adversarial attack on
  dnn-based black-box object detectors,'' {\em Journal of Network and Computer
  Applications}, vol.~161, p.~102634, 2020.

\bibitem{wang_deep_2018}
Z.~Wang, ``Deep {Learning}-{Based} {Intrusion} {Detection} {With}
  {Adversaries},'' {\em IEEE Access}, vol.~6, pp.~38367--38384, 2018.

\bibitem{clements_rallying_2019}
J.~H. Clements, Y.~Yang, A.~Sharma, H.~Hu, and Y.~Lao, ``Rallying {Adversarial}
  {Techniques} against {Deep} {Learning} for {Network} {Security},'' {\em
  CoRR}, vol.~abs/1903.11688, 2019.

\bibitem{ibitoye2019analyzing}
O.~Ibitoye, O.~Shafiq, and A.~Matrawy, ``Analyzing adversarial attacks against
  deep learning for intrusion detection in iot networks,'' {\em arXiv preprint
  arXiv:1905.05137}, 2019.

\bibitem{piplai2020nattack}
A.~Piplai, S.~S.~L. Chukkapalli, and A.~Joshi, ``Nattack! adversarial attacks
  to bypass a gan based classifier trained to detect network intrusion,'' {\em
  arXiv preprint arXiv:2002.08527}, 2020.

\bibitem{lin_idsgan:_2018}
Z.~Lin, Y.~Shi, and Z.~Xue, ``Idsgan: {Generative} adversarial networks for
  attack generation against intrusion detection,'' {\em arXiv preprint
  arXiv:1809.02077}, 2018.

\bibitem{peng2019adversarial}
X.~Peng, W.~Huang, and Z.~Shi, ``Adversarial attack against dos intrusion
  detection: An improved boundary-based method,'' in {\em 2019 IEEE 31st
  International Conference on Tools with Artificial Intelligence (ICTAI)},
  pp.~1288--1295, IEEE, 2019.

\bibitem{apruzzese2019evaluating}
G.~Apruzzese, M.~Colajanni, and M.~Marchetti, ``Evaluating the effectiveness of
  adversarial attacks against botnet detectors,'' in {\em 2019 IEEE 18th
  International Symposium on Network Computing and Applications (NCA)},
  pp.~1--8, IEEE, 2019.

\bibitem{umer2017flow}
M.~F. Umer, M.~Sher, and Y.~Bi, ``Flow-based intrusion detection: Techniques
  and challenges,'' {\em Computers \& Security}, vol.~70, pp.~238--254, 2017.

\bibitem{davis_data_2011}
J.~J. Davis and A.~J. Clark, ``Data preprocessing for anomaly based network
  intrusion detection: {A} review,'' {\em computers \& security}, vol.~30,
  no.~6-7, pp.~353--375, 2011.

\bibitem{draper-gil_characterization_2016}
G.~Draper-Gil, A.~H. Lashkari, M.~S.~I. Mamun, and A.~A. Ghorbani,
  ``Characterization of encrypted and vpn traffic using time-related,'' in {\em
  Proceedings of the 2nd international conference on information systems
  security and privacy ({ICISSP})}, pp.~407--414, 2016.

\bibitem{tavallaee_detailed_2009}
M.~Tavallaee, E.~Bagheri, W.~Lu, and A.~A. Ghorbani, ``A detailed analysis of
  the {KDD} {CUP} 99 data set,'' in {\em 2009 {IEEE} {Symposium} on
  {Computational} {Intelligence} for {Security} and {Defense} {Applications}},
  pp.~1--6, July 2009.

\bibitem{goodfellow_generative_2014}
I.~Goodfellow, J.~Pouget-Abadie, M.~Mirza, B.~Xu, D.~Warde-Farley, S.~Ozair,
  A.~Courville, and Y.~Bengio, ``Generative adversarial nets,'' in {\em
  Advances in {Neural} {Information} {Processing} {Systems} ({NIPS})},
  pp.~2672--2680, 2014.

\bibitem{kennedy_particle_2010}
J.~Kennedy, ``Particle swarm optimization,'' {\em Encyclopedia of machine
  learning}, pp.~760--766, 2010.

\bibitem{guyon_introduction_2003}
I.~Guyon and A.~Elisseeff, ``An introduction to variable and feature
  selection,'' {\em Journal of machine learning research}, vol.~3, no.~Mar,
  pp.~1157--1182, 2003.

\bibitem{sharafaldin_toward_2018}
I.~Sharafaldin, A.~H. Lashkari, and A.~A. Ghorbani, ``Toward {Generating} a
  {New} {Intrusion} {Detection} {Dataset} and {Intrusion} {Traffic}
  {Characterization}.,'' in {\em Proceedings of the 2nd international
  conference on information systems security and privacy ({ICISSP})},
  pp.~108--116, 2018.

\bibitem{qayyum2005taxonomy}
A.~Qayyum, M.~Islam, and M.~Jamil, ``Taxonomy of statistical based anomaly
  detection techniques for intrusion detection,'' in {\em Proceedings of the
  IEEE Symposium on Emerging Technologies, 2005.}, pp.~270--276, IEEE, 2005.

\bibitem{sperotto2010overview}
A.~Sperotto, G.~Schaffrath, R.~Sadre, C.~Morariu, A.~Pras, and B.~Stiller, ``An
  overview of ip flow-based intrusion detection,'' {\em IEEE communications
  surveys \& tutorials}, vol.~12, no.~3, pp.~343--356, 2010.

\bibitem{papernot2017practical}
N.~Papernot, P.~McDaniel, I.~Goodfellow, S.~Jha, Z.~B. Celik, and A.~Swami,
  ``Practical black-box attacks against machine learning,'' in {\em Proceedings
  of the 2017 ACM on Asia conference on computer and communications security},
  pp.~506--519, 2017.

\bibitem{kerckhoffs1883cryptographic}
A.~Kerckhoffs, ``La cryptographic militaire,'' {\em Journal des sciences
  militaires}, pp.~5--38, 1883.

\bibitem{eberhart_comparing_2000}
R.~C. Eberhart and Y.~Shi, ``Comparing inertia weights and constriction factors
  in particle swarm optimization,'' in {\em Proceedings of the 2000 congress on
  evolutionary computation}, vol.~1, pp.~84--88, IEEE, 2000.

\end{thebibliography}


	
	\appendices	
	
	\section{Correlation Test between Overhead of Feature-space and Traffic-space}
	\label{corr}
	
	Intuitively, it is easy to understand that the more difference between the two series of traffic, the more difference in their features' value. In turn, we speculate that the farther malicious traffic needs to move in the feature-space, the greater its overhead of mutation. To prove that, we conduct the following experiment to test the correlation between distance in feature-space and mutation overhead in traffic-space. The distance in feature-space is defined as Euclidean distance of two feature vectors, while mutation overhead in traffic-space is defined as the summation of delay overhead and crafted traffic volume overhead, which correspond to overhead budget $l_{t}$ and $l_{c}$ in our experiments respectively (Note that, we used actual overhead without budget limitation in this test). For illustration purposes, we use the notations $\mathcal{O}_{t}(\cdot)$ and $\mathcal{O}_{c}(\cdot)$ to represent the time elapsed and  volume of a series of traffic, and still use  $\mathcal{L}(\cdot,\cdot)$ to represent the Euclidean distance between two feature vectors, and $\mathcal{E}(\cdot)$ to represent feature extraction.
	
	\textbf{Correlation Test}. The procedure of the correlation test is as follows:
	\begin{enumerate}[leftmargin=1.5em]
		\item We extract 10,000 features from the malicious traffic in {\small\emph{\textsf{Botnet}}} traffic set denoted with $\boldsymbol{t}$ and call them  \emph{malicious features} denoted with $\boldsymbol{f}_{mal}$.
		
		\item For each malicious feature, we randomly sample 10 \emph{targeted features} denoted with $\boldsymbol{f}_{tar}$ at a distance range of $[100,1000]$ (according to the distribution in our traffic sets) from the malicious feature. Thus, we have 100,000 sample points in total, each one consists of a \emph{malicious feature} $\boldsymbol{f}_{mal}$ and a \emph{targeted feature} $\boldsymbol{f}_{tar}$.
		
		\item We execute the proposed PSO-based mutation algorithm (Alg. 2) for each $\boldsymbol{f}_{mal}$ and use its corresponding $\boldsymbol{f}_{tar}$ as the guide for traffic mutation (similar to adversarial feature in our attack). The mutated traffic solved after execution of the algorithm is denoted with $\hat{\boldsymbol{t}}$.
		
		\item We keep \emph{valid} sample points---$\mathcal{E}(\hat{\boldsymbol{t}})$ successfully approached the $\boldsymbol{f}_{tar}$  through the algorithm---for further test. Here we preserve sample points that satisfy $1-\frac{\mathcal{L}(\mathcal{E}(\hat{\boldsymbol{t}}),\boldsymbol{f}_{tar})}{\mathcal{L}(\boldsymbol{f}_{mal},\boldsymbol{f}_{tar})}>0.7$.
		
		\item For preserved sample points, we test the correlation  between feature-space distance $\mathcal{L}(\boldsymbol{f}_{mal},\boldsymbol{f}_{tar})$ and traffic-space overhead $(\frac{\mathcal{O}_{t}(\hat{\boldsymbol{t}})}{\mathcal{O}_{t}(\boldsymbol{t})} + \frac{\mathcal{O}_{c}(\hat{\boldsymbol{t}})}{\mathcal{O}_{c}(\boldsymbol{t})})$.
	\end{enumerate}
	
	\textbf{Results and conclusion.} Finally, we compute the Pearson Correlation Coefficient (PCC) of the preserved sample points, and the result is $\textbf{0.8664}$ (strong correlation).
	Consequently, we draw the conclusion that adversarial features can efficiently reduce the overhead of mutating traffic by saving the distance of movement in feature-space.

	\section{High-level Feature Summarization and Basic Mutation Operators}
	\label{app1}
	As mentioned in Section \ref{sec5.2}, we propose a high-level summarization of features used in ML-based NIDSs, and then design basic mutation operators to affect high-level features. In this appendix, we demonstrate the versatility of our summarization method and how basic mutation operators affect all high-level features by using existing NIDSs as case studies.
	
	First, widely-used feature extraction in existing work \cite{umer2017flow,davis_data_2011,mirsky_kitsune:_2018,draper-gil_characterization_2016,tavallaee_detailed_2009} can be generally divided into three steps:
	\begin{enumerate}[leftmargin=1.5em]
		\item \textbf{Choice of data form}. Network traffic is generally processed in two forms:
		\emph{packets} and \emph{sessions} (or flows). The difference between them is session-based extractions look at aggregated information of related packets.
		
		\item \textbf{Choice of basic measurements}. Three widely-used measurements are \emph{size} (e.g., packets' length), \emph{count} (e.g.,  \# packets), and \emph{time} (e.g., inter-arrival time between packets). 
		
		\item \textbf{Process of basic measurements}. Given measurements in certain data forms, existing extraction methods prefer window-based collection (e.g., data with same source IP or within a fixed time interval) and/or statistics computing (e.g., mean and variance) to get advanced features.
	\end{enumerate}

	\begin{figure}[]
		\centering
		\includegraphics[width=\linewidth]{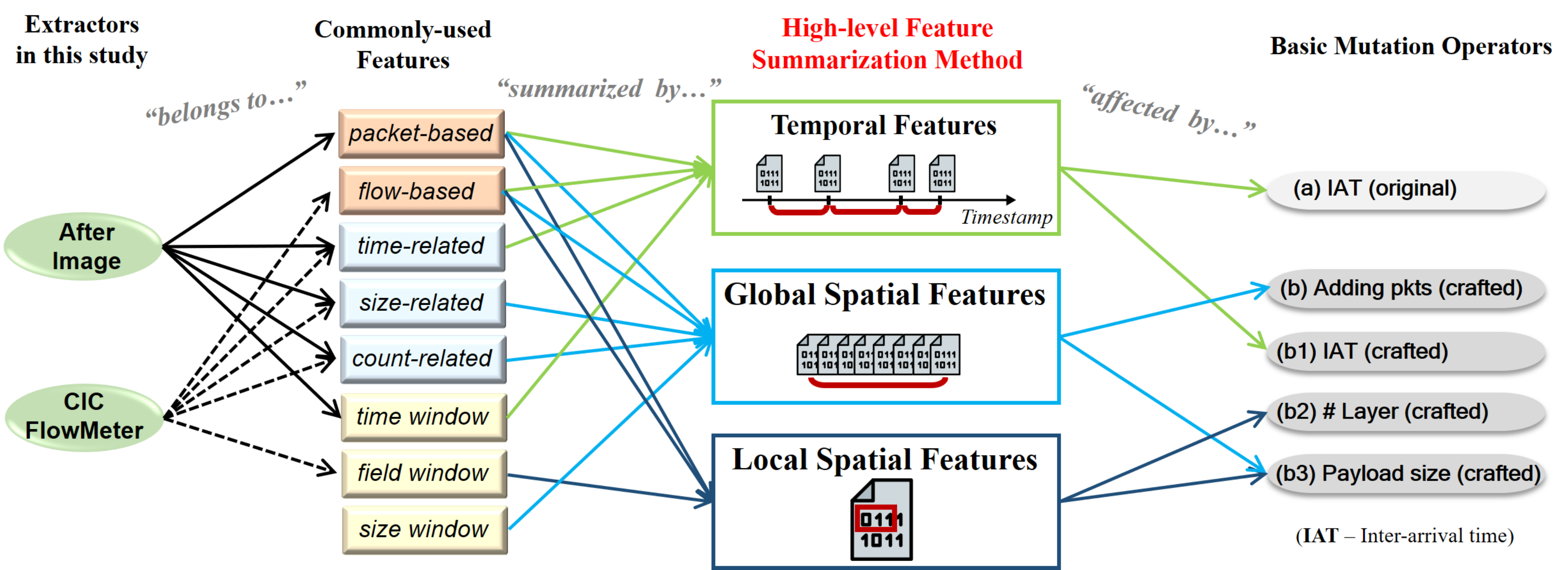}
		\caption{\small Illustration of the versatility of our summarization method and how basic mutation operators affect all high-level features}
		\label{figapp}
	\end{figure}
	
	Then in Fig. \ref{figapp}, we show how two extractors used in this work and commonly-used features be summarized by our summarization method, and how mutation operators affect all features.

	\section{PSO-based Traffic Mutation Algorithm}
	\label{app3}
	
	The PSO-based traffic mutation algorithm is shown in Algorithm \ref{alg1}. In each iteration, we firstly evaluate each particle's evasive effectiveness (on line 6) and update individual best and global best positions (on lines 7-8), which are respectively used to compute cognitive force (on line 11) and social force (on line 12). Then, each particle's $\boldsymbol{v}$ is updated by multiplying constant weights $\omega$, $c_{1}$, and $c_{2}$ with inertia, cognitive, and social items, respectively (on line 14). Each particle's $\boldsymbol{x}$ is then updated according to $\boldsymbol{v}$ (on line 15).
	
	\begin{algorithm}[!t]
		\small
		\setstretch{0.9}
		\SetKwComment{Comment}{$\triangleright$\ }{}
		\SetKwComment{Commentld}{$\triangledown$\ }{} 
		\SetFuncSty{asciifamily}
		\SetCommentSty{em}
		\SetKwFunction{Rebuild}{Rebuild}
		\SetKwFunction{Initialize}{Initialize}
		\SetKwFunction{UpdateX}{UpdateX}
		\SetKwFunction{Vectorize}{Vectorize}
		
		\caption{PSO-based Traffic Mutation Algorithm}
		\label{alg1}
		
		\KwIn{Hyperparameters $\omega, c_{1}, c_{2}, N_{iter}, N_{swarm}$ in PSO;\\
			\ \ attacker's overhead budget $l_{c}, l_{t}$;\\
			\ \ the original malicious traffic $\mathsf{T}$;\\
		}
		\KwOut{The mutated (/evasive) malicious traffic $\hat{\mathsf{T}}$. }
		
		\For{each $\boldsymbol{t}$ in $\mathsf{T}$} 
		{
			$\boldsymbol{t}_{v}\leftarrow$\Vectorize{$\boldsymbol{t}$}.\Comment*[f]{meta-info vectorization}
			
			$\mathsf{S}\leftarrow$\Initialize{$\boldsymbol{t}_{v},N_{swarm},l_{c}, l_{t}$}.\Comment*[f]{\small{initialize population}}
			
			\For{$i = 1$ to $N_{iter}$}
			{
				\For{$j = 1$ to $N_{swarm}$}
				{
					
					$\mathsf{S}.\boldsymbol{d}_{j} \leftarrow\mathcal{L}\big(
					\mathcal{E}'($\Rebuild{$\mathsf{S}.\boldsymbol{x}_{j}$}$),
					\{\mathbf{F}_{adver},\mathbf{F}_{ben}\}\big)$\;
					
					$\mathsf{S}.\boldsymbol{y}_{j}\leftarrow \rm{max}(\mathsf{S}.\boldsymbol{y}_{j},\mathsf{S}.\boldsymbol{d}_{j}) $\;
					
					$\mathsf{S}.\hat{\boldsymbol{y}}\leftarrow \rm{max}(\mathsf{S}.\hat{\boldsymbol{y}},\mathsf{S}.\boldsymbol{y}_{j}) $\;
					
				}
				\For(\Comment*[f]{update each $\boldsymbol{v}$ and $\boldsymbol{x}$}){$j = 1$ to $N_{swarm}$}
				{
					$\boldsymbol{v}_{cog} \leftarrow \mathsf{S}.\boldsymbol{y}_{j} -\mathsf{S}.\boldsymbol{x}_{j}$\Comment*{cognitive force}
					
					$\boldsymbol{v}_{soc} \leftarrow \mathsf{S}.\hat{\boldsymbol{y}} -\mathsf{S}.\boldsymbol{x}_{j}$\Comment*{social force}
					
					Randomly sample $r_{1},r_{2} \sim p_{\operatorname{uniform}(0,1)}$.
					
					$\mathsf{S}.\boldsymbol{v}_{j} \leftarrow \omega \mathsf{S}.\boldsymbol{v}_{j} + r_{1}c_{1}\boldsymbol{v}_{cog} + r_{2}c_{2}\boldsymbol{v}_{soc}$\;
					
					$\mathsf{S}.\boldsymbol{x}_{j}\leftarrow$ \UpdateX{$\mathsf{S}.\boldsymbol{x}_{j},\mathsf{S}.\boldsymbol{v}_{j},l_{c}, l_{t}$}.
				}
			}
			Append $\hat{\boldsymbol{t}}\leftarrow$
			\Rebuild{$\mathsf{S}.\hat{\boldsymbol{y}}$} into $\hat{\mathsf{T}}$.
		}
		\Return{$\hat{\mathsf{T}}$}
	\end{algorithm}
	
	\section{Adversarial Feature Evaluation Algorithm}
	\label{app21}
	Algorithm \ref{alg2} shows the specific robustness evaluation method mentioned in Section \ref{sec6}. we proactively simulate the evasion attack and measure the MMR (on line 7). Then by considering whether this feature vector can evade the classifier, a penalty or reward is added to adversarial robustness score (on lines 8-9). Finally, the adversarial robustness of a feature set is quantized into the score between $-1$ to $1$ (on line 12) of each feature.

	\begin{algorithm}[]
		\small
		\setstretch{0.9}
		\SetKwComment{Comment}{$\triangleright$\ }{}
		\SetKwComment{Commentld}{$\triangledown$\ }{} 
		\SetFuncSty{asciifamily}
		\SetCommentSty{em}
		\SetKwFunction{MMR}{MMR}
		
		\caption{Adversarial Feature Evaluation Algorithm}
		\label{alg2}
		\KwIn{$\mathbf{F}_{mal}$, $\hat{\mathbf{F}}_{mal}$, $\mathbf{F}_{adver}$, $\mathbf{F}_{ben}$, anomaly threshold $h$;}
		\KwOut{Adversarial feature score $\boldsymbol{s}$ of each dimension.}
		
		$n_{d} \leftarrow$ the dimensionality of a feature vector\;
		
		$n_{f} \leftarrow$ Number of features in $\mathbf{F}_{mal}$ or $\hat{\mathbf{F}}_{mal}$\;
		
		Initialize $\boldsymbol{s}$ with $n_{d}$ zeros.
		
		\For{each $\boldsymbol{f}$,$\hat{\boldsymbol{f}}$ in $\mathbf{F}_{mal}$,$\hat{\mathbf{F}}_{mal};$ $i = 0$ to $n_{f}-1$} 
		{
			Initialize an $\boldsymbol{r}$ with $n_{d}$ zeros.
			
			\For(\Comment*[f]{each dimension}){$j = 1$ to $n_{d}$}
			{
				$\boldsymbol{r}_{j} \leftarrow$\MMR{$\boldsymbol{f}_{j},\hat{\boldsymbol{f}}_{j},\mathbf{F}_{adver}, \mathbf{F}_{ben}$}\;
				
				\Commentld{add a penalty if successfully evading}
				\lIf{$\mathcal{E}(\boldsymbol{f})>h$ and $\mathcal{E}(\hat{\boldsymbol{f}})<h$}
				{
					$\boldsymbol{s}_{j} \leftarrow \boldsymbol{s}_{j} - \boldsymbol{r}_{j}$
				}
				\lElse(\Comment*[f]{add a reward}){
					$\boldsymbol{s}_{j} \leftarrow \boldsymbol{s}_{j} + (1-\boldsymbol{r}_{j})$
				}
				
			}
			
		}
		Normalize $\boldsymbol{s}$ through dividing each dimension by $n_{f}$.
		
		\Return{$\boldsymbol{s}$}
		
	\end{algorithm}
	
	\section{Verifying the malicious functionality}
	\label{app22}
	Detailed results of verifying malicious functionality of all six attack traffic sets are listed in Table \ref{tab21}.
	Note that, although the \emph{attack effect} cannot be measured in some cases, in fact the results of \emph{attack effect} are generally the same as \emph{malicious behavior}. For example, in {\small\emph{\textsf{Brute Force}}}, we do not know the true password of the victim server, but if we guarantee that all the original password attempts exist in the mutated traffic, then obviously the final result is the same.
	
	\begin{table}[]
		\caption {Comparison of the malicious functionality}
		\renewcommand\arraystretch{0.85}
		\centering
		\footnotesize
		\begin{minipage}[t]{\linewidth}
			\centering
			\vspace{-1.25mm}
			\centerline{(a) \emph{\textsf{Botnet}}}
			\vspace{0.15mm}
			\begin{tabular}{c|c|c|c}
				\hline
				\textbf{Indicators} & \textbf{Original} & \textbf{Mutated} & \textbf{Comparison} \\
				\hline
				\begin{tabular}[c]{@{}c@{}}Number of open \\ devices scanned\end{tabular}& 8 & 8 & --- \\ 
				\cdashline{1-4}[1.75pt/2pt]
				\begin{tabular}[c]{@{}c@{}}Total number of scans\end{tabular}& 8500 & 8500 & --- \\ 
				\cdashline{1-4}[1.75pt/2pt]
				\begin{tabular}[c]{@{}c@{}}Time elapsed\end{tabular}& 0.795s & 2.364s & $\uparrow$ (197\%) \\ 
				\hline
			\end{tabular}
		\end{minipage}
		
		\begin{minipage}[t]{\linewidth}
			\vspace{1.5mm}
			\centering
			\centerline{(b) \emph{\textsf{Fuzzing}}}
			\vspace{0.1mm}
			\begin{tabular}{c|c|c|c}
				\hline
				\textbf{Indicators} & \textbf{Original} & \textbf{Mutated} & \textbf{Comparison} \\
				\hline
				\begin{tabular}[c]{@{}c@{}}Impact of fuzzing\\on target systems\end{tabular}&\multicolumn{3}{c}{\begin{tabular}[c]{@{}c@{}}\scriptsize{This cannot be simulated because we have no} \\ \scriptsize{specific information about the targeted system}\end{tabular}} \\ 
				\cdashline{1-4}[1.75pt/2pt]
				\begin{tabular}[c]{@{}c@{}}\# pkts containing\\fuzzing payload\end{tabular}& 5353 & 5353 & --- \\ 
				\cdashline{1-4}[1.75pt/2pt]
				\begin{tabular}[c]{@{}c@{}}Time elapsed\end{tabular}& 3.81s & 2.90s & $\downarrow$ (24\%) \\ 
				\hline
			\end{tabular}
		\end{minipage}
		
		\begin{minipage}[t]{\linewidth}
			\vspace{1.5mm}
			\centering
			\centerline{(c) \emph{\textsf{SSDP DoS}}}
			\vspace{0.1mm}
			\begin{tabular}{@{}c@{}|c|c|c}
				\hline
				\textbf{Indicators} & \textbf{Original} & \textbf{Mutated} & \textbf{Comparison} \\
				\hline
				\begin{tabular}[c]{@{}c@{}}Impact of DoS attack\\on targeted systems\end{tabular}&
				\multicolumn{3}{c}{\begin{tabular}[c]{@{}c@{}}\scriptsize{This cannot be simulated because we have no} \\ \scriptsize{specific information about the targeted system}\end{tabular}} \\ 
				\cdashline{1-4}[1.75pt/2pt]
				\begin{tabular}[c]{@{}c@{}}Attack bandwidth\end{tabular}& 35Mbps & 20Mbps & $\downarrow$ (39\%) \\ 
				\cdashline{1-4}[1.75pt/2pt]
				\begin{tabular}[c]{@{}c@{}}Time elapsed\end{tabular}& 
				\multicolumn{3}{c}{\begin{tabular}[c]{@{}c@{}}\scriptsize{This can be reflected by the prevent} \\ \scriptsize{indicator (bandwidth) }\end{tabular}} \\ 
				\hline
			\end{tabular}
		\end{minipage}
		
		\begin{minipage}[t]{\linewidth}
			\vspace{1.5mm}
			\centering
			\centerline{(d) \emph{\textsf{Brute Force}}}
			\vspace{0.1mm}
			\begin{tabular}{@{}c@{}|c|c|c}
				\hline
				\textbf{Indicators} & \textbf{Original} & \textbf{Mutated} & \textbf{Comparison} \\
				\hline
				\begin{tabular}[c]{@{}c@{}}Whether the targeted\\FTP server is cracked\end{tabular}&                                                                                                                                             
				\multicolumn{3}{c}{\begin{tabular}[c]{@{}c@{}}\scriptsize{This cannot be simulated because we do not} \\ \scriptsize{know the true password of the targeted system}\end{tabular}} \\ 
				\cdashline{1-4}[1.75pt/2pt]
				\begin{tabular}[c]{@{}c@{}}Total number of\\password attempts\end{tabular}& 60 & 60 & --- \\ 
				\cdashline{1-4}[1.75pt/2pt]
				\begin{tabular}[c]{@{}c@{}}Time elapsed\end{tabular}& 2.83s & 11.26s & $\uparrow$ (298\%) \\ 
				\hline
			\end{tabular}
		\end{minipage}
		
		\begin{minipage}[t]{\linewidth}
			\vspace{1.5mm}
			\centering
			\centerline{(e) \emph{\textsf{Port Scan}}}
			\vspace{0.1mm}
			\begin{tabular}{c|c|c|c}
				\hline
				\textbf{Indicators} & \textbf{Original} & \textbf{Mutated} & \textbf{Comparison} \\
				\hline
				\begin{tabular}[c]{@{}c@{}}Number of open \\ ports scanned\end{tabular}& 3 & 3 & --- \\ 
				\cdashline{1-4}[1.75pt/2pt]
				\begin{tabular}[c]{@{}c@{}}Total number of \\ scans\end{tabular}& 4810 & 4810 & --- \\ 
				\cdashline{1-4}[1.75pt/2pt]
				\begin{tabular}[c]{@{}c@{}}Time elapsed\end{tabular}& 6.55s & 26.89s & $\uparrow$ (310\%) \\ 
				\hline
			\end{tabular}
		\end{minipage}
		
		\begin{minipage}[t]{\linewidth}
			\vspace{1.5mm}
			\centering
			\centerline{(f) \emph{\textsf{DDoS}}}
			\vspace{0.1mm}
			\begin{tabular}{@{}c@{}|c|c|c}
				\hline
				\textbf{Indicators} & \textbf{Original} & \textbf{Mutated} & \textbf{Comparison} \\
				\hline
				\begin{tabular}[c]{@{}c@{}}Impact of DDoS attack\\on targeted systems\end{tabular}&
				\multicolumn{3}{c}{\begin{tabular}[c]{@{}c@{}}\scriptsize{This cannot be simulated because we have no} \\ \scriptsize{specific information about the targeted system}\end{tabular}} \\ 
				\cdashline{1-4}[1.75pt/2pt]
				\begin{tabular}[c]{@{}c@{}}Attack bandwidth\end{tabular}& 107Mbps & 68Mbps & $\downarrow$ (36\%) \\ 
				\cdashline{1-4}[1.75pt/2pt]
				\begin{tabular}[c]{@{}c@{}}Time elapsed\end{tabular}& 
				\multicolumn{3}{c}{\begin{tabular}[c]{@{}c@{}}\scriptsize{This can be reflected by the prevent} \\ \scriptsize{indicator (bandwidth) }\end{tabular}} \\ 
				\hline
			\end{tabular}
		\end{minipage}
		
		\label{tab21}
	\end{table}
	
\end{document}